\documentclass[aps,amsmath,amssymb,reprint]{revtex4-1}
\usepackage{graphicx}
\usepackage{dcolumn}
\usepackage{bm}
\usepackage[utf8]{inputenc}
\usepackage{natbib}
\usepackage[T1]{fontenc}
\usepackage{mathptmx}
\usepackage{etoolbox}
\makeatletter 
\def\@email#1#2{%
 \endgroup
 \patchcmd{\titleblock@produce}
  {\frontmatter@RRAPformat}
  {\frontmatter@RRAPformat{\produce@RRAP{*#1\href{mailto:#2}{#2}}}\frontmatter@RRAPformat}
  {}{}
}%
\usepackage{hyperref}
\hypersetup{colorlinks,linkcolor={blue},citecolor={blue},urlcolor={blue}}
\begin{document}
\preprint{AIP/123-QED}
\title {Performance of a Brownian information engine through potential profiling: Optimum output requisites, Heating-to-Refrigeration transition and their Re-entrance }
\author{Rafna Rafeek}
\affiliation{Department of Chemistry and Center for Molecular and Optical Sciences \& Technologies, Indian Institute of Technology Tirupati, Yerpedu 517619, Andhra Pradesh, India}
\author{Debasish Mondal}
  \email{debasish@iittp.ac.in}
\affiliation{Department of Chemistry and Center for Molecular and Optical Sciences \& Technologies, Indian Institute of Technology Tirupati, Yerpedu 517619, Andhra Pradesh, India}
\date{\today}
\begin{abstract}
Brownian Information engine (BIE) harnesses the energy from a fluctuating environment by utilizing the associated information change in the presence of a single heat bath. The engine operates in a space-dependent confining potential and requires an appropriate feedback control mechanism. 
In general, the feedback controller has three different steps: measurement, feedback, and relaxation. The feedback step is related to a sudden change in the potential energy that is essential for a nonzero work output. 
BIE utilises the amount of information (surprise) acquired during the measurement step for the energy output. However, due to the relaxation process, a certain amount of acquired information is lost or becomes unavailable. So, controlling information loss during relaxation is crucial for the overall efficiency of the engine. The net (available) information, therefore, can be monitored by tuning the feedback controller and the shape of the confining potential. 
In this paper, we explore the effect of the shape modulation of the confining potential, which may have multiple stable valleys and unstable hills, on the net available information and, hence, the performance of a BIE that operates under an asymmetric feedback protocol.  We examine the optimal performance requirements of the BIE and the amount of maximum work output under different potential profiling. For monostable trapping, a concave shape in confining potential results in a higher work output than a convex one. We also find that hills and valleys in the confining potential may lead to multiple good operating conditions. An appropriate shape modulation can create a heater-refrigerator transition and their reentrance due to non-trivial changes in information loss during the relaxation process.
\end{abstract}
\maketitle
\section{INTRODUCTION}
The attempts to extract work from a single heat bath date back to the thought experiment of Maxwell's demon \cite{rex2017maxwell,maxwell2001theory, Szilard1929zphys, Landaueribm1961}. Maxwell's demon is a feedback controller which observes gas molecules within a single heat bath and uses the gathered information to extract work \cite{seifert2005, seifert2008, Sagwa2008prl, Sagawa2010prl, Sagawa2009prl,  Sagawa2012pre, Horowitz2010pre, Horowitz2011epl}, apparently violating the Second law of thermodynamics \cite{rex2017maxwell,maxwell2001theory}. However, the paradox was resolved in the seminal works of Szilard, Landauer, Bennett and others by considering the thermodynamic cost associated with the processing of the acquired information \cite{maxwell2001theory, Szilard1929zphys, Landaueribm1961, rex2017maxwell}. The resolution of the paradox revealed the connection between stochastic thermodynamics and information theory, particularly how information-theoretic measures like mutual information are connected to the entropy \cite{seifert2005, seifert2008, Sagawa2010prl, Sagwa2008prl, Sagawa2012pre, Sagawa2009prl, Horowitz2010pre, Horowitz2011epl}. The Jarzynski relation \cite{Jarzynski1997prl}, and subsequent advancements in fluctuation theorems \cite{seifert2005, jarzynski2011, Seifert_2012, Gaspard2004, das2012shape} set up the foundation of thermodynamic relations in a stochastic environment. In modern days, with the help of stochastic thermodynamics, one can explicitly incorporate information as a means to extract work and its connection to the free-energy difference between nonequilibrium states \cite{seifert2008, Sagawa2010prl, Sagwa2008prl, Sagawa2012pre, Sagawa2009prl, Horowitz2010pre, Horowitz2011epl}. In this spirit, one can employ appropriate cyclic feedback mechanisms, to extract work from a single heat bath by utilizing the mutual information gained during the measurement \cite{Abreu2011epl, Abreu2012prl, Bauer_2012, Mandal_2013, Taichi_2013, Pal2014pre,  Ashida2014pre, ali2022geometric, rafeek2023geometric}. It is thus possible to design a Brownian Information Engine (BIE) considering overdamped Brownian particles as working substance \cite{ Abreu2012prl, Bauer_2012, Mandal_2013, Taichi_2013, Pal2014pre,  Ashida2014pre, ali2022geometric, Abreu2011epl, Park2016pre, rafeek2023geometric, saha2021, rafeek2024, Paneru_2022}. The quantitative relation between the free energy, work done and the information change related to a state change (between two equilibrium states) in a BIE reads as \cite{seifert2005, seifert2008, Sagwa2008prl, Sagawa2010prl, Sagawa2009prl}:
   $ -\langle W \rangle \leq -\Delta F + k_BT (\langle I \rangle - \langle I_u \rangle)$,
here $\langle .. \rangle$ denotes the ensemble averaging. The inequality suggests that work extraction $(-\langle W \rangle)$ from a single heat bath (with temperature $T$) is bounded by the sum of free energy change $(\langle \Delta F \rangle)$  and the net available information $(\langle I \rangle - \langle I_u \rangle)$. We will define and describe the total information measured $(\langle I \rangle)$ and the unavailable information $(\langle I_u \rangle)$ later.

\paragraph*{} The intrinsic fluctuations in living systems are often rectified to produce different functional biophysical activities \cite{fang2019, Horowitz2013, Heiner_2021, borsley_2021,chakraborty2023, Tomé2018, Schmiedl2007, wu2014, xiao2009, Fritz2020, Horowitz2015, Mou1986}. The emerging concept of utilizing information as fuel to biological motors, enabling the directed motion and improved efficiencies of systems \cite{fang2019, borsley_2021, chakraborty2023}, has escalated the interest in understanding the information-energy exchange within the fluctuating environment \cite{barato2013, barato2014, Horowitz2013}. Consequently, the use of measurement information to facilitate the extraction of work from a single thermal bath has been studied primarily theoretically in both classical systems \cite{Abreu2011epl, Abreu2012prl, Bauer_2012,  Taichi_2013, Pal2014pre,  Ashida2014pre, ali2022geometric,  Mandal_2013, Park2016pre, rafeek2023geometric} and quantum systems \cite{Kim2011prl, Bruschi2015pre, Goold2016jphysA, koski2014pnas}. With the advancements in nonequilibrium statistical mechanics and significant progress in experimental techniques, feedback mechanisms utilizing information from Brownian systems have been extensively explored \cite{Abreu2011epl, Abreu2012prl, Bauer_2012,  Taichi_2013, Pal2014pre,  Ashida2014pre, ali2022geometric, Park2016pre, Mandal_2013, rafeek2023geometric, Paneru2020natcommun, Toyabe2010natphys, Berut2012nat, Paneru2018prl, Lopez2008prl, dago2021, dago2022}. Some of the important theoretical outcomes have recently been corroborated (verified) by experimental findings \cite{Toyabe2010natphys, Berut2012nat, Paneru2018prl, Paneru2020natcommun, archambault2024, Lopez2008prl, dago2021, dago2022} as well. The commonly realized BIE utilizes the positional information of the overdamped particles trapped in a harmonic potential \cite{Abreu2011epl, Abreu2012prl, Bauer_2012, Mandal_2013, Taichi_2013, Pal2014pre,  Ashida2014pre, Park2016pre, Toyabe2010natphys, Berut2012nat, Paneru2018prl, Paneru2020natcommun}. The feedback controller operates based on the information related to the measurement outcome and modifies the location of the potential centre. A sudden change in the potential energy of the trapped particle leads to work extraction in such a fluctuating environment. One approach to implementing such a feedback mechanism involves the instantaneous shift of the potential centre to a new feedback location $(x_f)$ if the measured particle position surpasses the specified measurement distance $(x_m)$. When the energy of the particle is lowered (on average) following the shift of the potential centre, indicating work extraction, the engine functions as a heater.
On the other hand, with an increase in the particle's potential energy, the engine functions as a refrigerator \cite{Paneru2020natcommun, Park2016pre, rafeek2023geometric}. Consequently, the details of the feedback strategy, the choice of the feedback parameters ($(x_m)$ and $(x_f)$), and the shape of the confining potential are expected to influence the engine's performance. 
 
\begin{figure}[!htb]
{\includegraphics[width=0.45\textwidth]{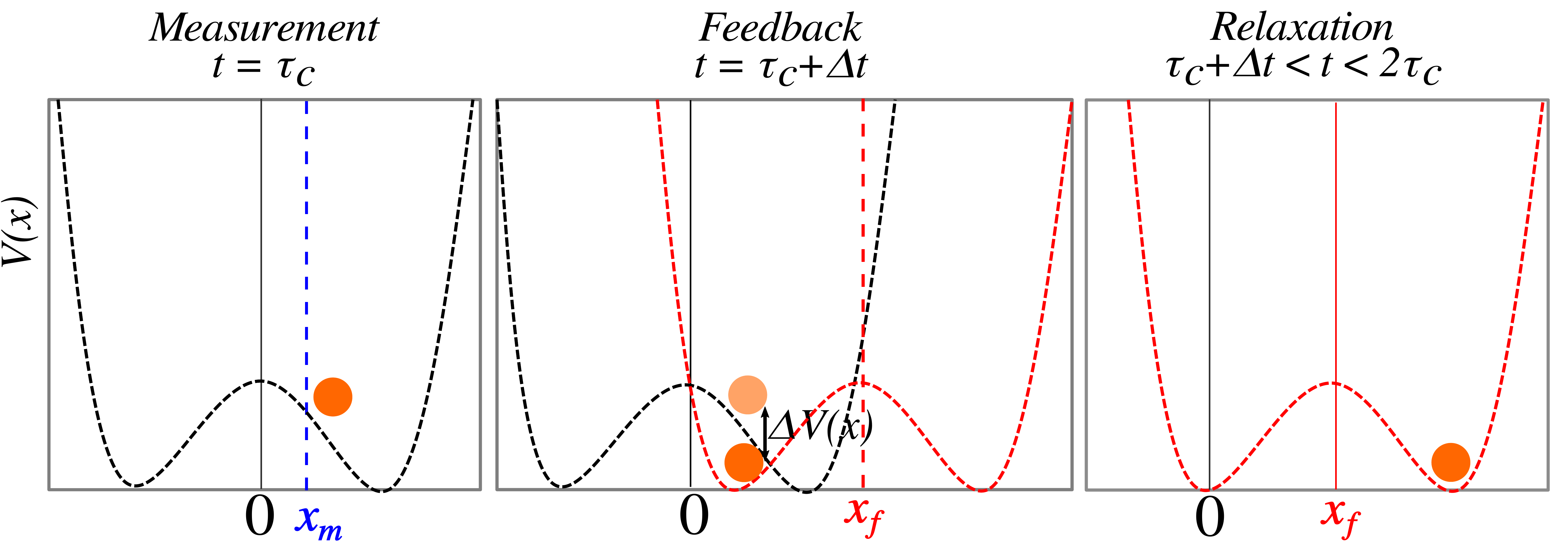}} 
     \caption{Schematic illustration of an asymmetric feedback cycle during $\tau_c \leq t \leq 2\tau_c$. The three-step feedback protocol comprises the following: (a) Measurement: At $t=\tau_c$, the confining potential is centred at zero ($\lambda = 0$). We measure the position ($x$) of the particle. (b) Feedback: From the measurement outcome, we estimate whether $x \ge x_m$ or not. $x_m$ is the measurement distance (blue-coloured dashed line).  If $x \ge x_m$, we shift the confinement centre instantaneously to the new feedback location ($\lambda = x_f$, red coloured dashed line), $\lambda = x_f$. Otherwise ($x < x_m$), the potential centre remains unchanged $\lambda = 0$. (c) Relaxation: The Particle is allowed to be relaxed with an unaltered potential centre until the next cycle begins.}
    \label{f2}
\end{figure}

\paragraph*{}Feedback is decided based on the measurement outcome. Two popular feedback strategies are frequently used in designing a BIE. When the potential shift is permitted in both directions, we refer to it as a symmetric protocol \cite{Ashida2014pre, ali2022geometric}; if restricted to a single direction, we mention it as an asymmetric protocol.  In this paper, we consider a commonly used asymmetric feedback controller \cite{Abreu2011epl, Abreu2012prl, rafeek2023geometric, Paneru2018prl, paneru2018pre, Park2016pre}. One measurement distance ($x_m$) is chosen externally and arbitrarily. The potential centre is shifted to a feedback site based on the measurement outcome (if the particle's position exceeds $x_m$) \cite{Paneru2018prl, paneru2018pre, Park2016pre}. 
The measurement predictability and thermal equilibration of the system depend on the dispersion $\sigma$ of the particle's position. The best operating requisites of work extraction for such BIE are as follows: measurement distance $x_m \approx 0.6\sigma$ and feedback location $x_f = 2x_m$ \cite{Abreu2011epl, Abreu2012prl, Park2016pre, Paneru2018prl}. One can also find a condition of measurement distance and feedback location for a heater-to-refrigerator transition \cite{Park2016pre, Paneru2020natcommun}. In our recent study, a Geometric Brownian information engine subjected to a monostable entropic enclosure \cite{zwanzig1992, reguera2001, mondal2010diffusion, ghosh2010,  ali2024}, we identified the analogous conditions on $x_m$ and $x_f$ for optimal work extraction and the heater-to-refrigeration transition \cite{rafeek2023geometric}. The details of the feedback procedure will be discussed in the next section (Fig.~\ref{f2}).
\paragraph*{}
The  other important requirement for the information-energy exchange in a fluctuating environment is constraining the Brownian particle within a spatially varying external potential (or effective potential).
The external potential landscape determines the dispersion of the steady-state distribution of particle position and thus is expected to significantly impact the work harvestation from the rectified fluctuations.
Generally, a harmonic potential as a working confining technique is widespread because of its easy experimental implementation \cite{Paneru2018prl, archambault2024}. However, in the presence of complex potential landscapes, the noise-induced barrier crossing \cite{marie2020,zwanzig1988,mondal2011, ghosh2007,mondal2010entropic} and other different noise-assisted phenomena, for instance, stochastic resonance \cite{mcnamara1989, mondal2016resonance, zhu2022, mondal2012entropic}, resonant activation \cite{doering1992, reimann1995, jan1996, mondal2010resonant, chattoraj2014dynamics}, and ratchet rectification \cite{magnasco1993,  mondal2009roughratchet, mondal2016ratchet, kato2013quantum, wagoner2016}, exhibit substantial and intricate behaviours that showcase the constructive role of the underlying fluctuating environment. Thus, it is important to examine the effects of different potential profiles on energy harnessing from the fluctuating systems in an information engine. In particular, it will be interesting to explore the effect of confinement design on the criteria for best performance and heater-to-refrigeration transition of the information engine operating with a particular feedback strategy. In a recent study with a symmetric feedback controller, we reported that concave confinement is more efficient for an information-energy exchange for a Brownian information engine than convex trapping \cite{rafna_pot_2025}.\begin{figure}[!htb]
   {\includegraphics[width=0.33
\textwidth]{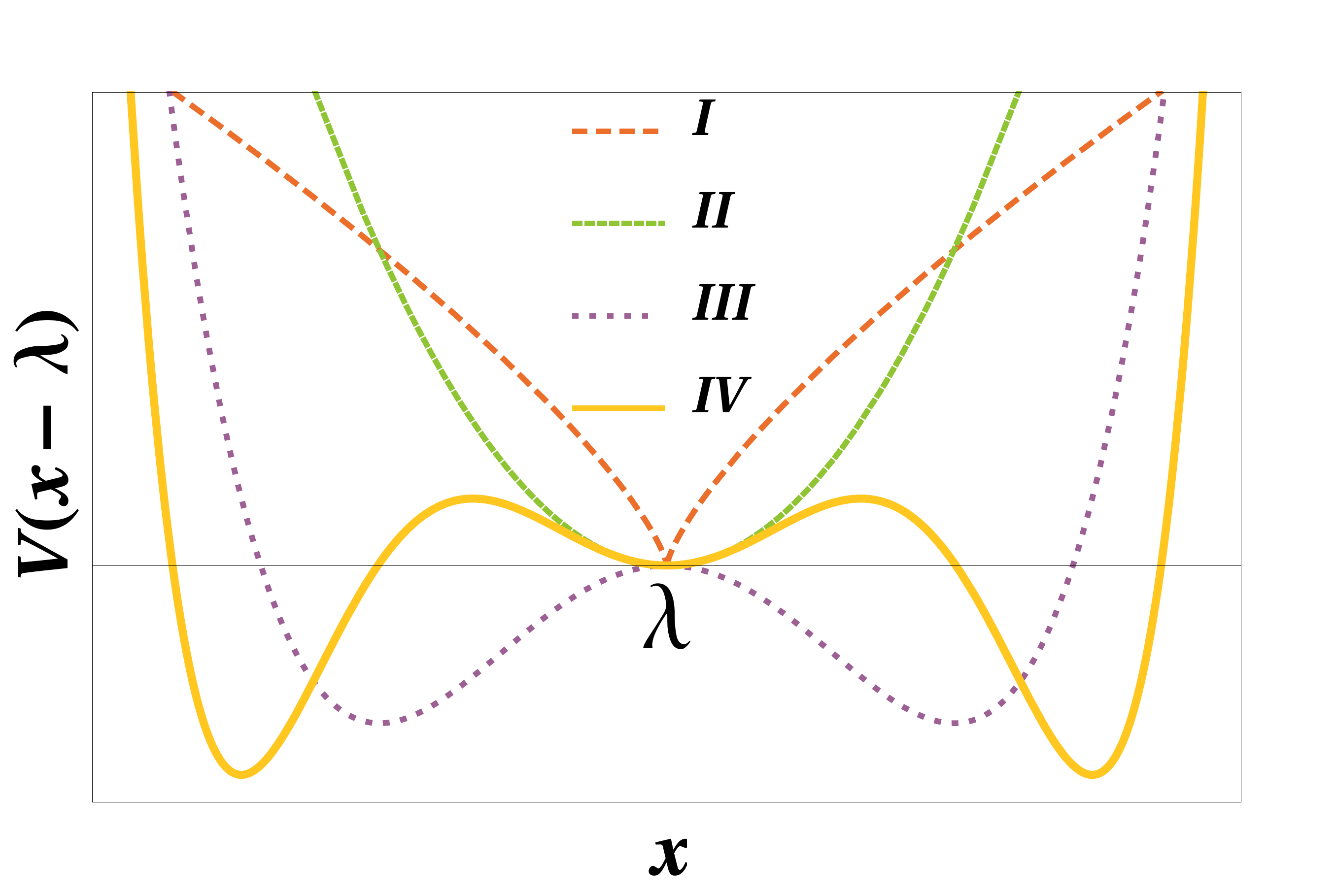}} 
     \caption{The schematic representation of centrosymmetric potential profiles used in this study: (I) a concave monostable potential, (II) a convex monostable potential, (III) bistable confinement, and (IV) a triple well potential trap. $\lambda$ indicates the spatial location of the potential centre.}
    \label{f1}
\end{figure}
\paragraph*{}
With these considerations, we pose the following questions on the dependence of the operating essentials of a BIE on the shape and nature of the confining potential: a) Can the increasing concavity and convexity of the single-well confinement influence the conditions for optimal performance? b) If a monostable potential provides a single optimal condition for work extraction, can a bistable or multistable potential offer more than one set of good choices of feedback conditions? c) Following the same spirit, can a potential landscape with multiple energy hills and valleys induce heater-to-refrigerator transitions more than once, possibly leading to a re-entrance phenomenon? To address the aforementioned issues, we focus on tuning the confining potential landscape in the presence of an asymmetric feedback cycle. We consider three distinct scenarios of potential landscapes: (i) a single-well potential with a stable potential centre where the shape (stiffness) is adjusted (varied), (ii) a continuous tuning of monostable trapping to bistable confinement with an unstable potential centre, and (iii)  a gradual introduction of multistability in potential trapping by a centrosymmetric single well to triple well crossover.
\section{Modelling the Information Engine}
\subsection{Confining potential and the Langevin equation of motion}

 \begin{figure*}[!htb]
\begin{minipage}[b]{0.29\linewidth}
\centering
 \includegraphics[width=1\textwidth]{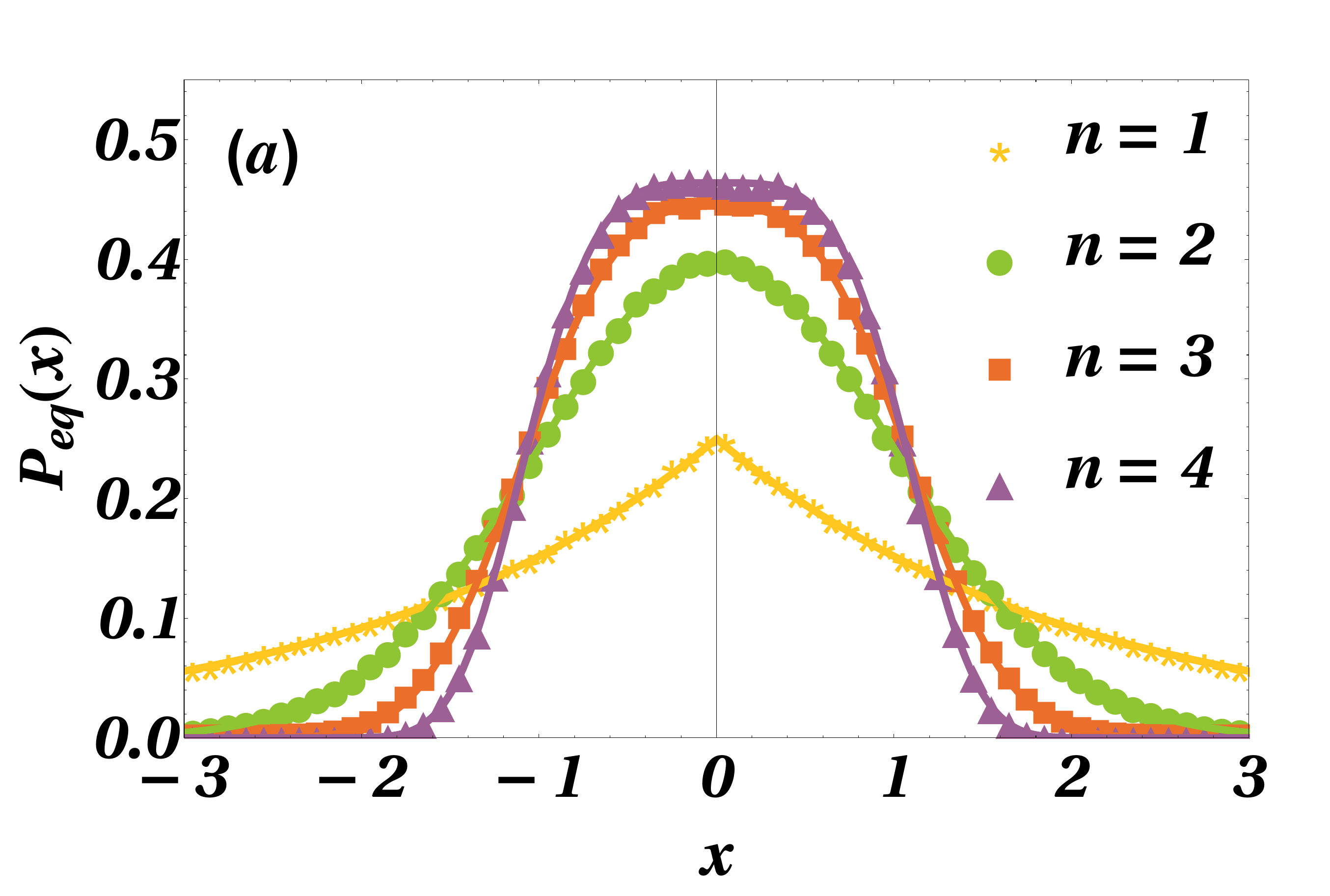}
\end{minipage}
\hspace{0.01cm}
\begin{minipage}[b]{0.29\linewidth}
\centering
\includegraphics[width=1\textwidth]{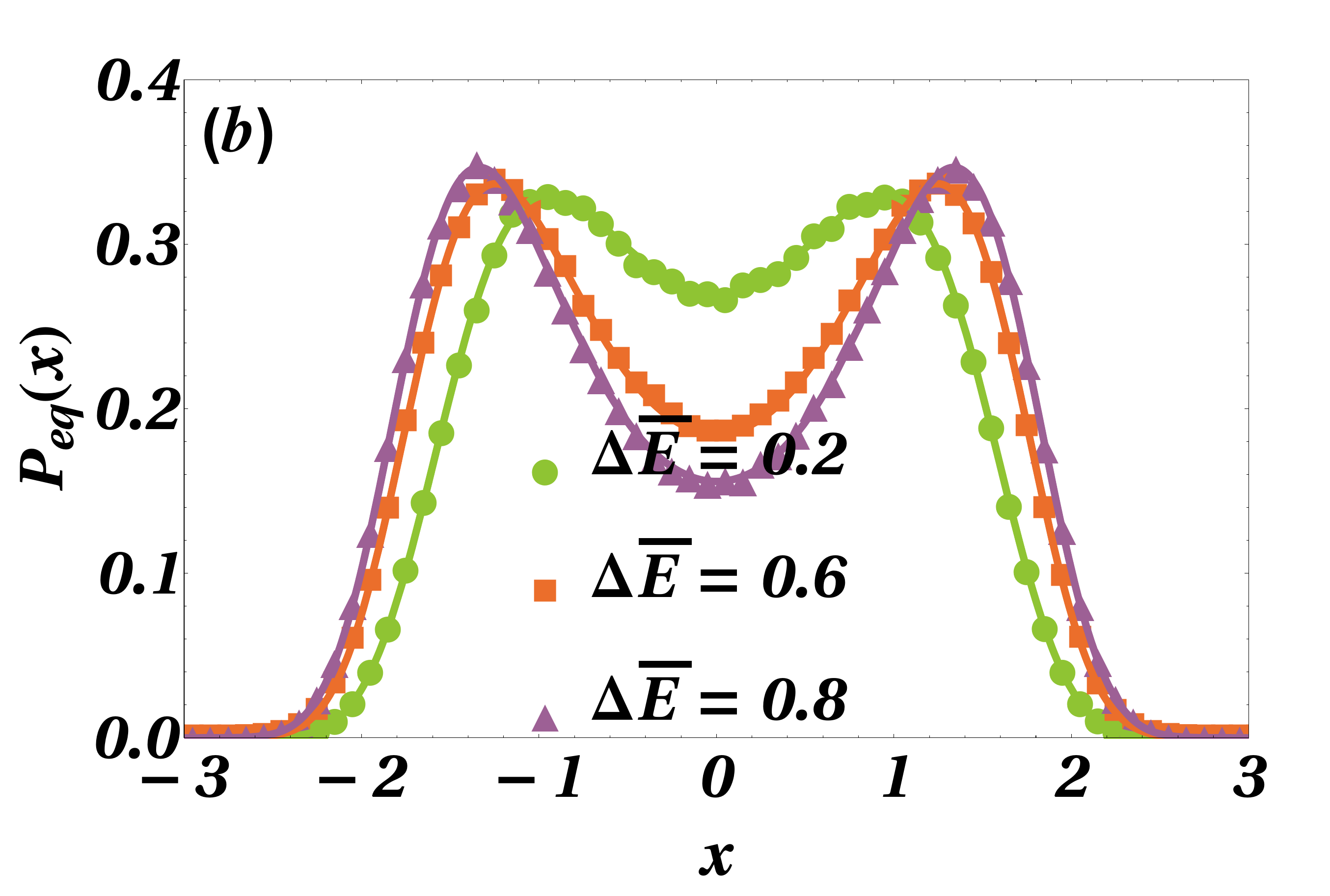}
\end{minipage}
\hspace{0.01cm}
\begin{minipage}[b]{0.29\linewidth}
\centering
\includegraphics[width=1\textwidth]{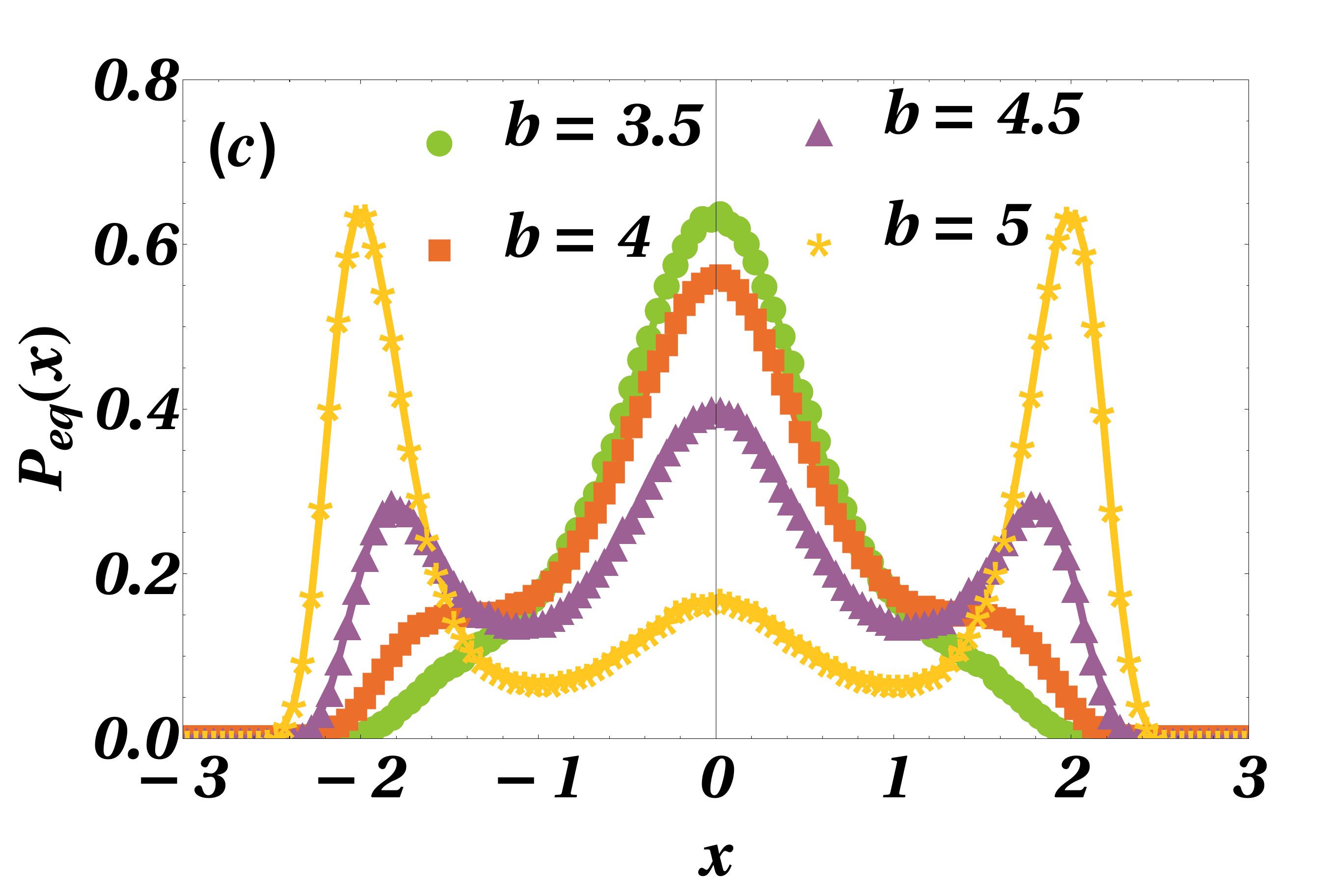}
\end{minipage}
\caption{The equilibrium probability distribution ($P_{eq}(x)$) for a particle confined in (a) mono-stable potential of form $V(x)=a|x|^n$ with different values of $n$ for $a=\frac{1}{2}$. 
(b) Bistable potential of form $V(x) = -\frac{a}{2}x^2 + \frac{b}{4}x^4$ with different values of scaled barrier height $\Delta \overline{E}$ $(\Delta E = a^2/4b)$ for $b = 1$. (c) Triple well potential of form $V(x) = \frac{a}{2}x^2 - \frac{b}{4}x^4 + \frac{c}{6}x^6$ with different values of $b$, with $a = 4$ and $c = 1$. In all cases, solid lines represent the theoretical predictions and points are obtained from numerical simulation (Eq.~\ref{2}). In all cases, $k_BT$ and $\gamma$ are chosen as unity. }
\label{f3}
\end{figure*}
We consider an overdamped Brownian particle confined in an external arbitrary centrosymmetric 1-D potential, $V(x-\lambda)$. Where $\lambda$ is a constant number and denotes the position of the potential centre.  We restrict the overall form of $V(x-\lambda)$ as monostable (both convex or concave around the potential centre), bistable or a triple-well potential in nature and tune their shape whenever required. We deliberately consider the shape of $V(x-\lambda)$ in a way that practically confines the particle's movement up to a certain (finite) distance from the potential centre. The schematics of the potential types used in this study are shown in Fig.~\ref{f1}. These simplified model potential shapes ensure an easy understanding of underlying physics but sufficiently capture different aspects related to the efficiency of a BIE.  The Langevin equation of non-interacting Brownian particle under consideration reads as:
\begin{eqnarray}\label{2}
\begin{aligned}
    \dot{x} &= -\frac{V^{\prime}(x-\lambda)}{\gamma} + \sqrt{2D} \eta(t),  \\
 \text{with} \;\; \left \langle \zeta (t)\right \rangle &= 0, \;\; \left \langle \zeta(t) \zeta ({t}')\right \rangle =\delta (t-{t}'),
\end{aligned}
 \end{eqnarray}
where $\gamma$ stands for friction coefficient, and $D$ corresponds to diffusion coefficient $(D=k_BT / \gamma)$,  $T$ is thermostat temperature, and  $k_B$ denotes the celebrated Boltzmann constant. Thermal fluctuations are modelled by a zero mean Gaussian white noise $\zeta(t)$. The prime notation in the $V^{\prime}(x-\lambda)$ is conventional and it represents the first derivative with respect to position $x$. Next, we consider a dimensionless description of the dynamical equation of motion (Eq.~\ref{2}).
We scale all relevant physical observables using a logical reference scale as reported in a recent experimental study on a Brownian information engine \cite{Paneru2018prl}. We scale an arbitrary variable $Y$ by a reference value $Y_r$, yielding dimensionless $\tilde Y$. We consider room temperature the reference temperature scale, $T_r \sim 293 K$. Thus, the corresponding thermal energy will be the energy reference scale, $ E_r = k_B T_r$ Joule. We set other relevant reference scales as follows: frictional coefficient as $\gamma_r \sim 18.8 \; nNm^{-1}s$, length scale as $x_r  \sim 20 \; nm$ and time $t_r = \sim 2 ms$ \cite{Paneru2018prl}. The reference unit of potential parameters can be obtained considering their dimensions. For instance, stiffness for the harmonic potential is $E_r / x_r^{2}$ i.e. $10 pN \; \mu m^{-1}$. However, we drop the tilde sign from the scaled observables in the rest of the manuscript for its simplicity.
\subsection{The asymmetric feedback protocol}
Initially (at $t=0$), we set the potential centre at the origin of the reference coordinate system $\lambda=0$ and allow the overdamped Brownian particle confined in an external potential landscape, $V(x - \lambda)$ to be equilibrated for a sufficiently long time. We define a reference measurement distance as $x_m$ and a feedback location as $(x_f)$ for operational purposes. Once the system reaches thermal equilibrium, the particle is subjected to the feedback cycle $(\text{at} \; t = \tau_c)$ as illustrated in Fig.~\ref{f2}. The feedback control comprises three sequential steps: measurement, feedback, and relaxation. At $t = \tau_c$, we measure the particle position $x$. Later in the feedback step, if the particle position is beyond the measurement distance $(x \geq x_m)$, the potential centre is instantaneously shifted to a new site, which we call feedback location $x_f$, $\text{i.e.}\; \lambda=x_f$. On the other hand, if the particle position doesn't cross the measurement distance $(x < x_m)$, the potential centre remains the same $(\lambda=0)$. During the relaxation step, the particle relaxes back to thermal equilibrium in the presence of an unaltered potential centre. 
The cycle is repeated to obtain a statistical average outcome of the observables. Typically, one needs to set a cycle time $(\tau_c)$ much larger than the thermal relaxation time  $(\tau_r)$ of the system, i.e. $\tau_c > \tau_r$. The feedback protocol is one of the commonly used controllers to devise a Brownian information engine for both experimental and theoretical studies \cite{Abreu2011epl, Ashida2014pre, Toyabe2010natphys, Lopez2008prl}. As explained earlier, the location of the measurement distance and feedback site are independent of each other and chosen externally.  The outcome performance ability of such an asymmetric controller thus depends on the choice of $x_m$ and $x_f$. 
\subsection{Calculating work, information  and beyond}
Next, we calculate the average work done due to the sudden change in potential energy during the feedback. As we consider the entire controller error-free, the instantaneous shift of the potential centre allows for the complete conversion of the change in the potential to extractable work. Therefore, the work extraction $(-W(x))$ related to a single feedback thus reads as:
\begin{equation}\label{3}
\begin{aligned}
   -W(x) & = V(x)-V(x-x_f), \;\;\; \text{if} \;\; x \geq x_m, &\\
    &= 0, \;\;\;\; \text{if} \;\; x < x_m.
\end{aligned}
\end{equation}
The feedback cycle is repeated and the average work extraction is written as:
\begin{equation}\label{4}
      -\langle W \rangle = -\int_{x_m}^{\infty} dxP_{eq}(x)W(x),
\end{equation}
Here, $P_{eq}(x)$ denotes the equilibrium probability distribution of particle position and can be obtained by solving the Fokker-Planck equation \cite{Risken, van1992stochastic} of motion (an alternative description of the Langevin dynamics, described in Eq.~\ref{2}) at long time as:
\begin{eqnarray} \label{5}
    \begin{aligned}
        P_{eq}(x) = \mathcal{N} \exp \bigg [ -\frac{V(x -\lambda ) }{k_B T} \bigg],
    \end{aligned}
\end{eqnarray}
where $\mathcal{N}$ is normalization constant. The $P_{eq}(x)$ distributions for three distinct types of potential profiling under consideration are shown in Fg.~\ref{f3}. The work done by the system is considered negative work, and the minus sign in front of $\langle W \rangle$ takes care of the sign consistency.

\paragraph*{}The notion of \textit{information} is related to the extent of uncertainty or surprisal related to the outcomes of a certain event.  The information linked to an event increases with a decrease in the probability of outcome. Consider an event outcome of $y$ with probability $p(y)$ in a range of zero to unity. The information associated with event outcome $y$ is defined as $I(y) = - \ln p(y)$.
A highly probable outcome, i.e. $(p(y) \to 1)$, will have lower surprisal. Whereas for less probable outcome $(p(y) \to 0)$, the associated surprisal diverges. In the present study, the normalised $P_{eq}(x)$ is a continuous observable, and its upper bound is not limited to unity. The net acquired information is equivalent to the Shannon entropy of the particle at its initial equilibrium. In this spirit, one can define a total average information acquired $ \langle I \rangle $ during the measurement step as \cite{Sagawa2010prl, Ashida2014pre, ali2022geometric}:
\begin{equation}\label{6}
    \begin{aligned}
       \langle I \rangle = -\int_{-\infty}^{\infty} dxP_{eq}(x)\ln[P_{eq}(x)].
    \end{aligned}
\end{equation}
During the relaxation process, some part of the information acquired during the measurement step is lost due to thermal relaxation. To compute the unavailable information, we consider the reverse feedback protocol. In this protocol, the particle is in equilibrium with the potential centred at the feedback location $(\lambda = x_f)$. We then abruptly shift the potential centre to origin $(\lambda = 0)$ regardless of the particle position and allow the system to be relaxed. The average unavailable information is given as \cite{Sagawa2010prl, Ashida2014pre, ali2022geometric}:
\begin{equation}\label{7}
       \langle I_u \rangle = -\int_{-\infty}^{x_m} dxP_{eq}(x)\ln[P_{eq}(x)] -\int_{x_m}^{\infty} dxP_{eq}(x)\ln[P_{eq}(x-x_f)].
\end{equation}
Analysing Eqs.~\ref{4}-\ref{7}, it can be shown that the extractable average work and the available information attain equality as: $-\langle W \rangle = k_BT ( \langle I \rangle -\langle I_u \rangle)$. The proposed error-free feedback cycle, thus, acts as a lossless information engine \cite{Paneru2018prl, ali2022geometric, rafeek2023geometric}. Also, one can verify the integral fluctuation theorem \cite{Sagawa2010prl, Sagawa2012pre, Ashida2014pre, Abreu2011epl} by substituting the work and information terms in the following relation:
\begin{equation}\label{8}
\begin{aligned}
      \langle \text{e}^{-(\overline{W}+I-I_u )} \rangle =
    \int_{-\infty}^{\infty} {dx} P_{eq}({x})\text{e}^{\overline{V}(x)-\overline{V}(0)}\frac{P_{eq}({x})}{P_{eq}(0)}
     \;\;= 1.
\end{aligned}
\end{equation}
Here onwards, the notation $\overline{z}$ is a scaled quantity and defined as $\overline{z}=\frac{z}{k_B T}$. Finally, we define the standard deviation of the particle position at an equilibrium state as $(\sigma)$ that can obtained from the following relation:
\begin{equation}\label{9}
    \begin{aligned}
     \sigma^2 = \int_{-\infty}^{\infty}x^2 P_{eq}(x)dx - \bigg(\int_{-\infty}^{\infty}x P_{eq}(x)dx \bigg)^2.
    \end{aligned}
\end{equation}
\subsection{Simulation details and experimental relevance}
Most of the results presented in this paper are either exact (analytical) or can be obtained theoretically using numerical integration. We use Eq.~\ref{5} to obtain the analytical expressions of $P_{eq} (x)$ for different potential set-up. We also numerically determine the equilibrium probability distribution using the overdamped Langevin equation (Eq.~\ref{2}) \cite{Hildebrandnumerical}.  Trajectories in order of $\sim 10^{7}$ are generated to compute the averages. To simulate trajectories of the particle's position, we have employed an improved Euler method with a time step of $10^{-3}$ units. The Gaussian noise is generated via the Box-Muller algorithm \cite{box1958}. When analytical integration is not feasible, numerical integration is carried out using Simpson's $1/3$ rule \cite{Hildebrandnumerical}.
 \begin{figure}
    \includegraphics[width=0.45\textwidth]{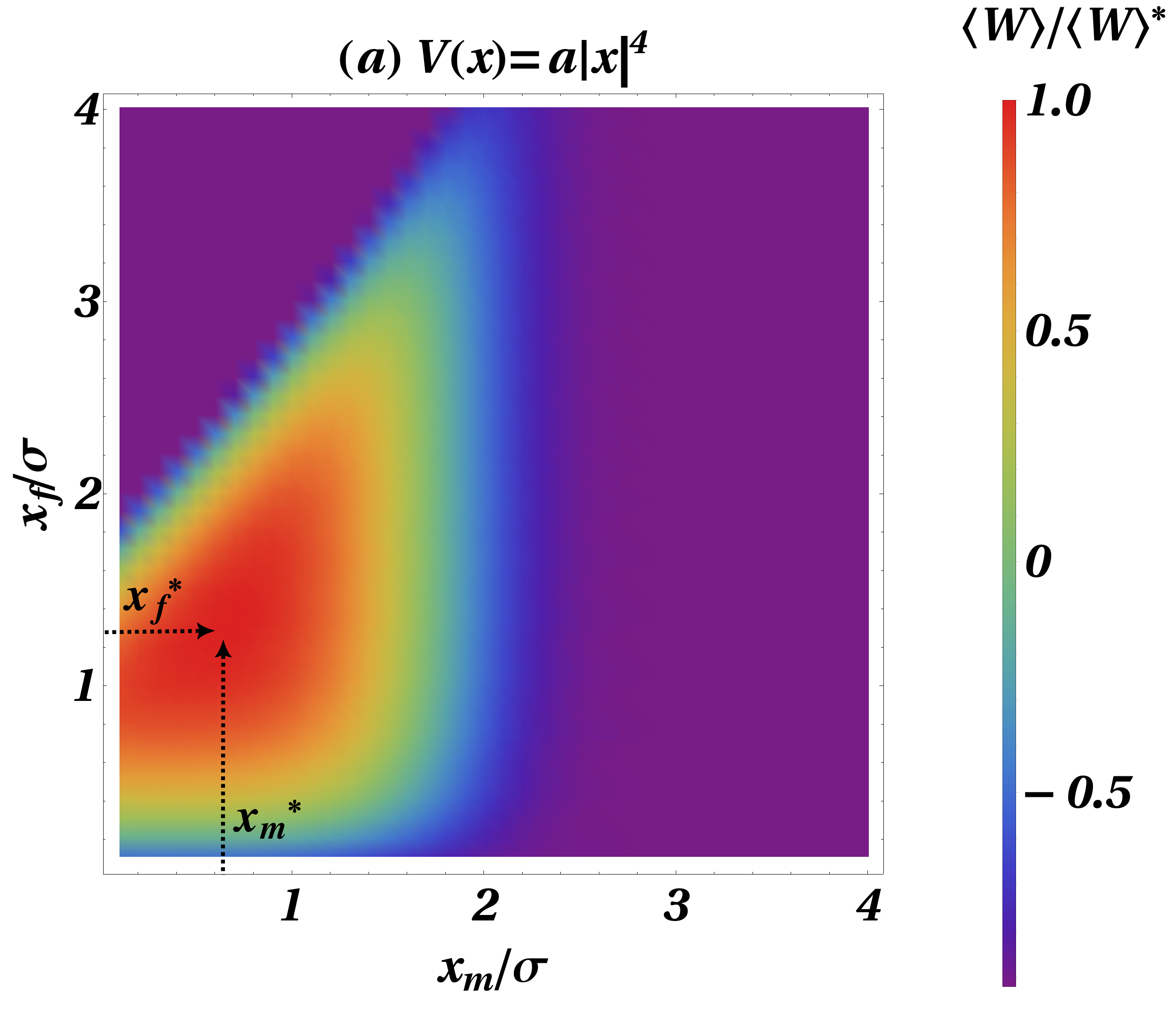}\\ \includegraphics[width=0.45\textwidth]{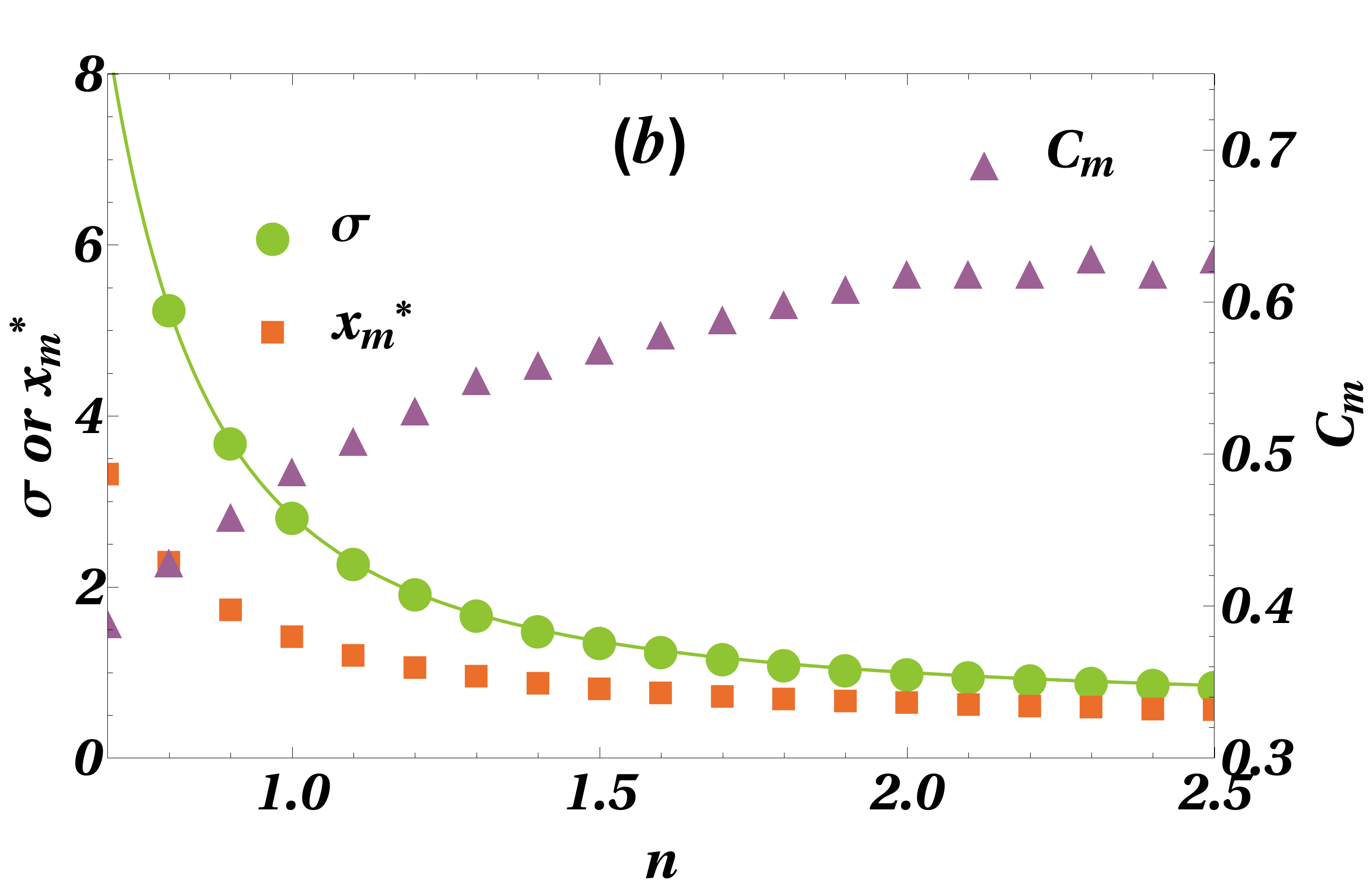}\caption{(a) The density plot of output worked ( scaled as $-\langle W \rangle/\langle W^* \rangle$) under different scaled feedback control parameters, $x_m$ and $x_f$ for single-well potential of form $V(x)=a |x|^4$. (b) Variation of standard deviation $(\sigma)$, optimal value of measurement distance $(x_m^*)$ and proportionality constant $(C_m)$, as a function of exponent $n$. The parameter set is chosen as: $a = 0.5$, $\gamma =1$ and $k_BT=1$.}   
    \label{f4}
\end{figure}
  As indicated earlier, different experimental attempts have been made to validate the theoretical findings of information thermodynamics \cite{Ramesh2025, Kim2021, Cao2022} of a BIE. Most of these experimental setups involve colloidal particles trapped by a focused laser beam, which creates harmonic potential and thus stimulates overdamped Brownian motion in a single well confinement \cite{Paneru2018prl}. In a different experimental set-up \cite{jop2008work} to simulate Brownian particle confined in a double well potential, one can use two focused laser beams to trap a polystyrene resulting in a quadratic potential. Sometime, electrostatic feedback can be employed to generate virtual double-well potentials acting on a micro-cantilever, which functions as an underdamped mechanical oscillator \cite{dago2021,dago2022, archambault2024}.
\section{RESULT AND DISCUSSIONS}
\subsection{Tuning the shape of a monostable confinement}
To begin with, we study a BIE operating within a single-well potential trap, described by $V(x)=a|x|^n$, where $a$ denotes the force constant and $n$ is a constant positive exponent, $n > 0$. $|x|$ refers to the absolute value of $x$. On increasing the value of the exponent, the potential shape changes from a concave $(n<1)$ to a convex $(n>1)$ one. The equilibrium distribution of particle position $(P_{eq}(x))$ and the standard deviation $(\sigma)$ (following Eqs.~\ref{5} and \ref{9}) read as:
\begin{equation}\label{10}
    \begin{aligned}
      P_{eq}(x) = \mathcal{N}_m  \exp  [- \overline{a}|x|^n  ], \; 
  \text{ and } \sigma^2 = \frac{\Gamma \left( \frac{3}{n} \right)}{a ^ {\frac{2}{n}} \Gamma \left( \frac{1}{n} \right) },
    \end{aligned}
\end{equation}

respectively. Where, $\mathcal{N}_m =   \frac{\overline{a}^{\frac{1}{n}}}{2 \Gamma (1+\frac{1}{n})}$, and $\Gamma(z)$ denotes a gamma function which is of the form $\Gamma(z) = \int_0^\infty t^{z-1} e^{-t} dt$. Fiq.~\ref{f3}(a) shows that the distribution ($P_{eq}(x)$) is symmetric and unimodal with varying degrees of tailedness, which depends on the extent of the concavity of the potential. 
As evident from the definition (Eqs.~\ref{3}-\ref{5} and Eq.~\ref{10}), the average work output of the engine $(-\langle W \rangle)$ depends on the choice of the exponent of the confining potential, measurement distance $(x_m)$ and feedback location $(x_f)$. We then estimate the average work obtained under different potential exponent with varying $x_m$ and $x_f$ to investigate the best performance requisites in such monostable constrain.
\begin{figure}[h!]
    \centering    
\includegraphics[width=0.46\textwidth]{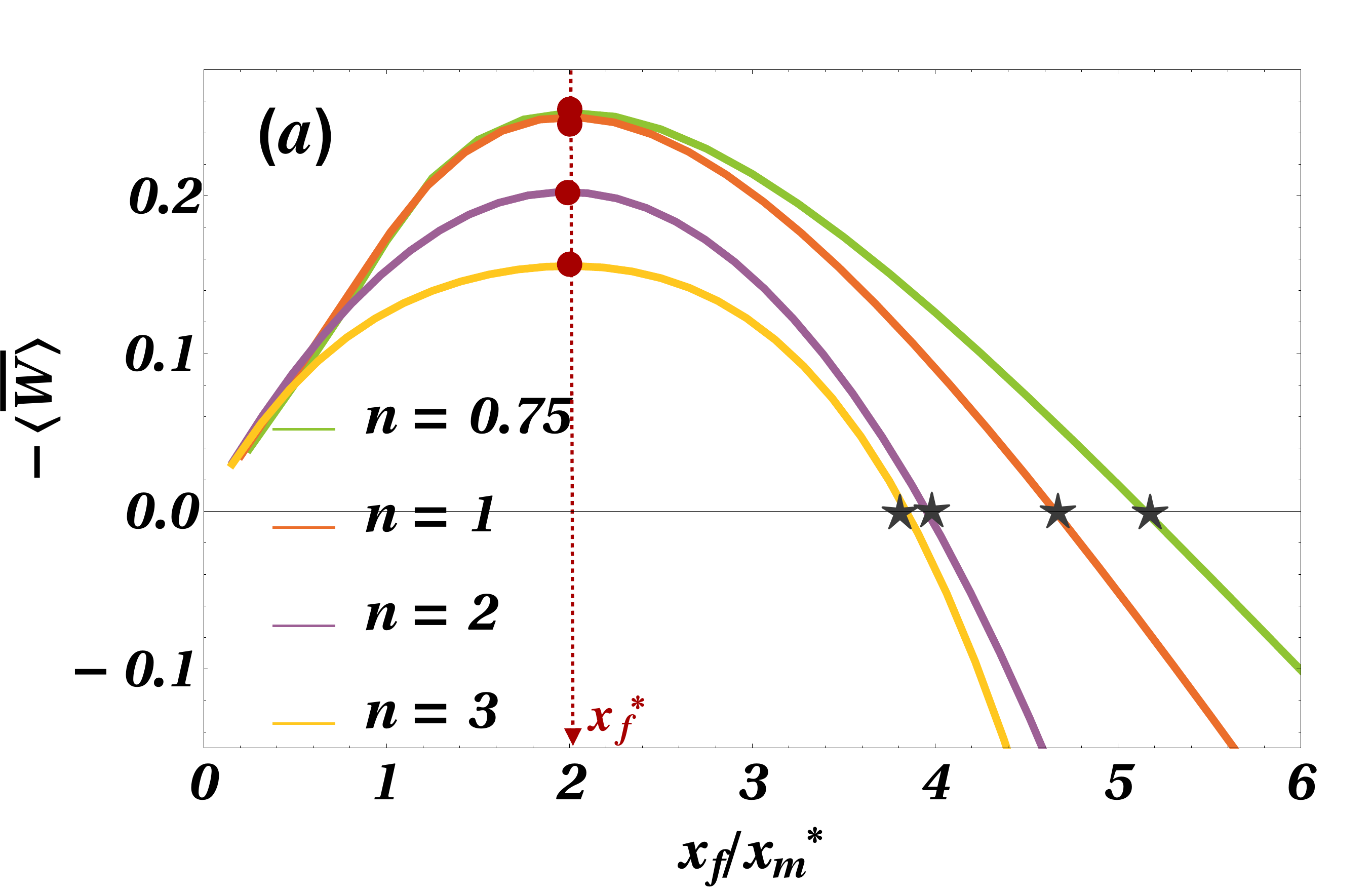}\\
\vspace{0.42cm}
\includegraphics[width=0.43\textwidth]{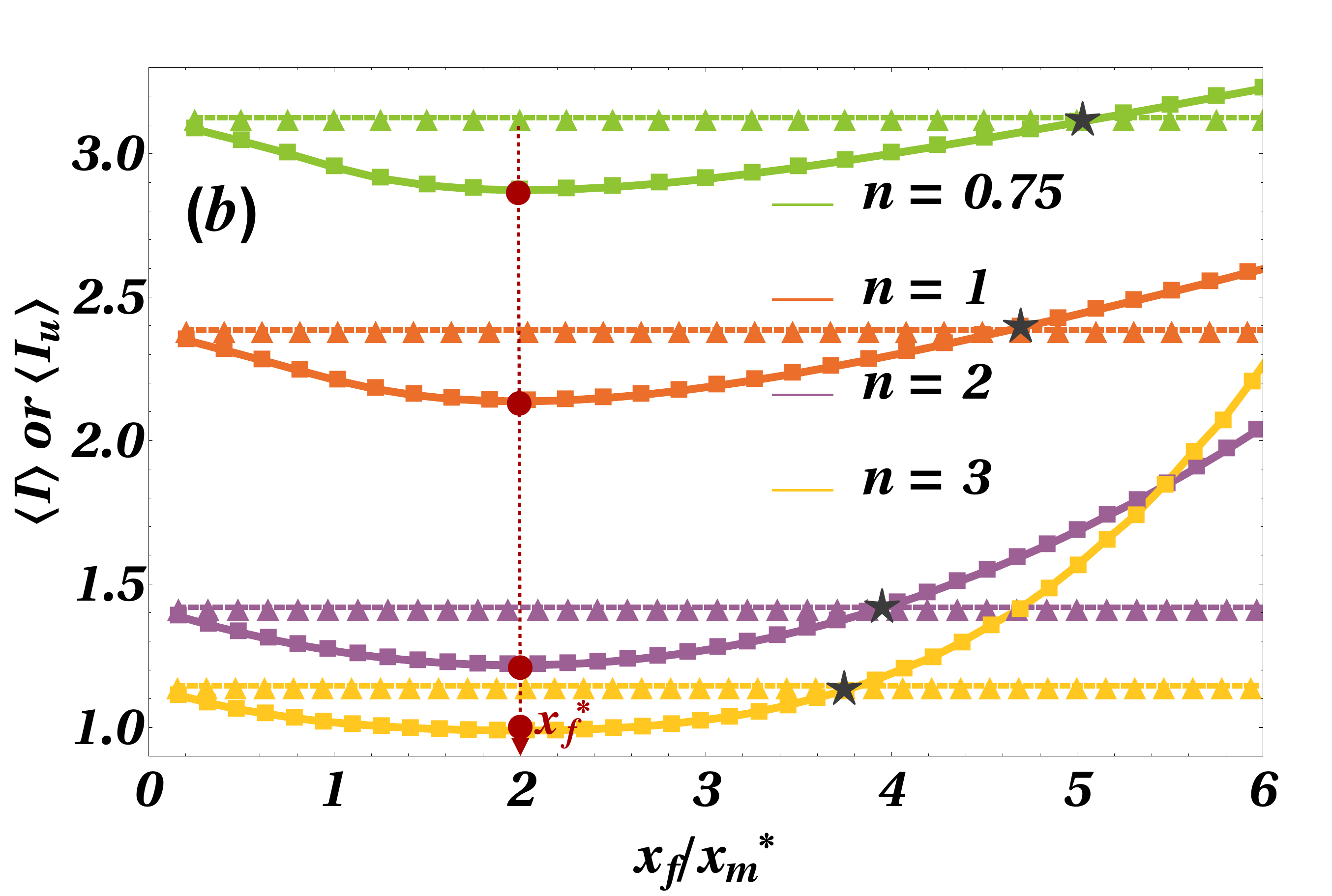}
    \caption{(a) Variation of a scaled work output $(-\langle \overline{W} \rangle)$, (b) average acquired information $(\langle I \rangle \text{, filled triangle-dotted lines })$ and average unavailable information $(\langle I_u \rangle  \text{, filled square-solid lines })$, as a function of scaled feedback location $(x_f/x_m^*)$ with measurement length $x_{m}^{*}$, for different values of power exponent $n$. The red-coloured filled circle indicates the best feedback locations $(x_f^*)$ that correspond to the best output $-\langle W \rangle^*$ (or least unavailable information), and the dark grey coloured filled star shows the heater-refrigerator transition point ($x_f^{inv}$). The parameter set is chosen as: $a = 0.5$, $\gamma =1$ and $k_BT=1$.}  
    \label{f5}
\end{figure}
Fig.~\ref{f4}(a) depicts a density plot of scaled work output from BIE with a potential exponent $(n = 4)$  in a 2-D space of $(x_m/ \sigma, \; x_f/\sigma)$. $\langle W^* \rangle$ is the extractable work under optimal feedback conditions. The observations are as follows. We find the optimal value of measurement and feedback site as $x_m^* \sim 0.65 \sigma$ and $x_f^* = 2x_m^*$. The density plot also reveals that the engine transitions from heater to refrigeration under certain combinations of $x_m$ and $x_f$. For instance, the choice of a very high measurement distance $(x_m \to high$) or a high feedback site $( x_f \to high)$ or both always function as refrigeration. In the limiting case of $x_f \to 0$, the feedback protocol is absent, leading to no work output $- \langle W \rangle = 0$ for obvious reasons. We perform a similar analysis of BIE under different potential exponent $(n > 0)$ (plots not shown). The study reveals the optimal requisites for maximum work extraction follow a general trend as $x_m^*= C_m \sigma$ and $x_f^*=2x_m^*$ for all positive values of $n$. The requisite of optimal choice of the feedback control parameter is, thus, general and consistent with the previous investigations on BIE with a harmonic oscillator as a working potential \cite{Park2016pre, Paneru2018prl}. However, it is important to note that the dispersion $\sigma$ decreases rapidly with rising $n$ (Fig~\ref{f4}(b)). In contrast, the proportionality constant $(C_m)$ increases slightly with increasing $n$ (Fig~\ref{f4}(b)). Therefore, the magnitude of the best measurement distance $x_m^*$ increases with the enhanced concavity of the potential (lower $n$). Next, we find the relative value in the maximum extractable output work $(-\langle W \rangle^*)$ under the best feedback controller $(x_m^*,\; x_f^*)$ for different power exponent $(n)$. For this purpose, we plot the variation of scaled work $-\langle \overline{W} \rangle$ as a function of scaled feedback location $(x_f/x_m^*)$ for different $n$ considering the corresponding $x_m^*$ as the measurement spot, see Fig~\ref{f5}(a). The output work shows a non-monotonic variation with increasing feedback location.  The magnitude of maximum work output $(-\langle \overline{W} \rangle ^*)$, as obtained at $x_f^* = 2 x_m^*$, increases with increasing concavity $(n < 1)$. A further increase of feedback location (higher than $x_f*$), the engine shows a transition from the heater to refrigeration beyond a specific feedback site $x_f=x_f^{inv}$. We find that the value of $x_f^{inv}$ increases with rising extent of concavity.
\paragraph*{}
To shine a light on the above-mentioned observations in terms of an information-energy exchange with varying $n$, we calculate the average acquired information $\langle I \rangle$ and average unavailable information $\langle I_u \rangle$, following Eqs.~\ref{6}-\ref{7} and \ref{10}. The Fig~\ref{f5}(b) depicts the variation of $\langle I \rangle$ and $\langle I_u \rangle$ as a function of the scaled feedback location $(x_f/x_m^*)$ under different potential shapes $(n)$ (with $x_m=x_m^*$). As per definition (Eq.~\ref{6}), the average acquired information for a given confining potential is independent of the choice of feedback site $x_f$ (Fig~\ref{f5}(b)). On the contrary, the unavailable information $(\langle I_u \rangle )$ shows a non-monotonic variation with scaled feedback location $(x_f/x_m^*)$, and it is minimum for a feedback location $x_f=2x_m^*$ (for all $n$). As the spread of $P_{eq}(x)$ becomes wider, both $\langle I \rangle $ and $\langle I_u \rangle$ increase with increasing concavity (decreasing $n$). 
However, the changes occur in such a way that the net available information $\langle I \rangle - \langle I_u \rangle$ at best $x_f^*$ increases proportionately with the extent of concavity of the confining potential.
This results in a higher maximum work output in concave trapping.
Increasing the feedback location further, the information lost during relaxation dominates the acquired information $(\langle I_u \rangle > \langle I \rangle )$, and the engine functions as a refrigerator. Again, the dominance of such unavailable information happens at a longer $x_f^{inv}$ for a concave potential compared to a convex confinement.
\subsection{Impact of an unstable potential center}
To comprehend the influence of potential with perturbed centre on work harvesting and functionality, we study the BIE with the confinement of form: $V(x)= -\frac{a}{2}x^2 + \frac{b}{4} x^4$, where $a$ and $b$ are potential parameters. By varying the parameter $a$, from zero to a non-zero positive value, the mono-stable centre of the potential can be perturbed, leading to a centrosymmetric bistable configuration with minima at $ x = \pm \sqrt{\frac{a}{b}}$ with a barrier top at the potential centre ($x=0$). 
\begin{figure}[!htb]
\centering
\includegraphics[width=0.45\textwidth]{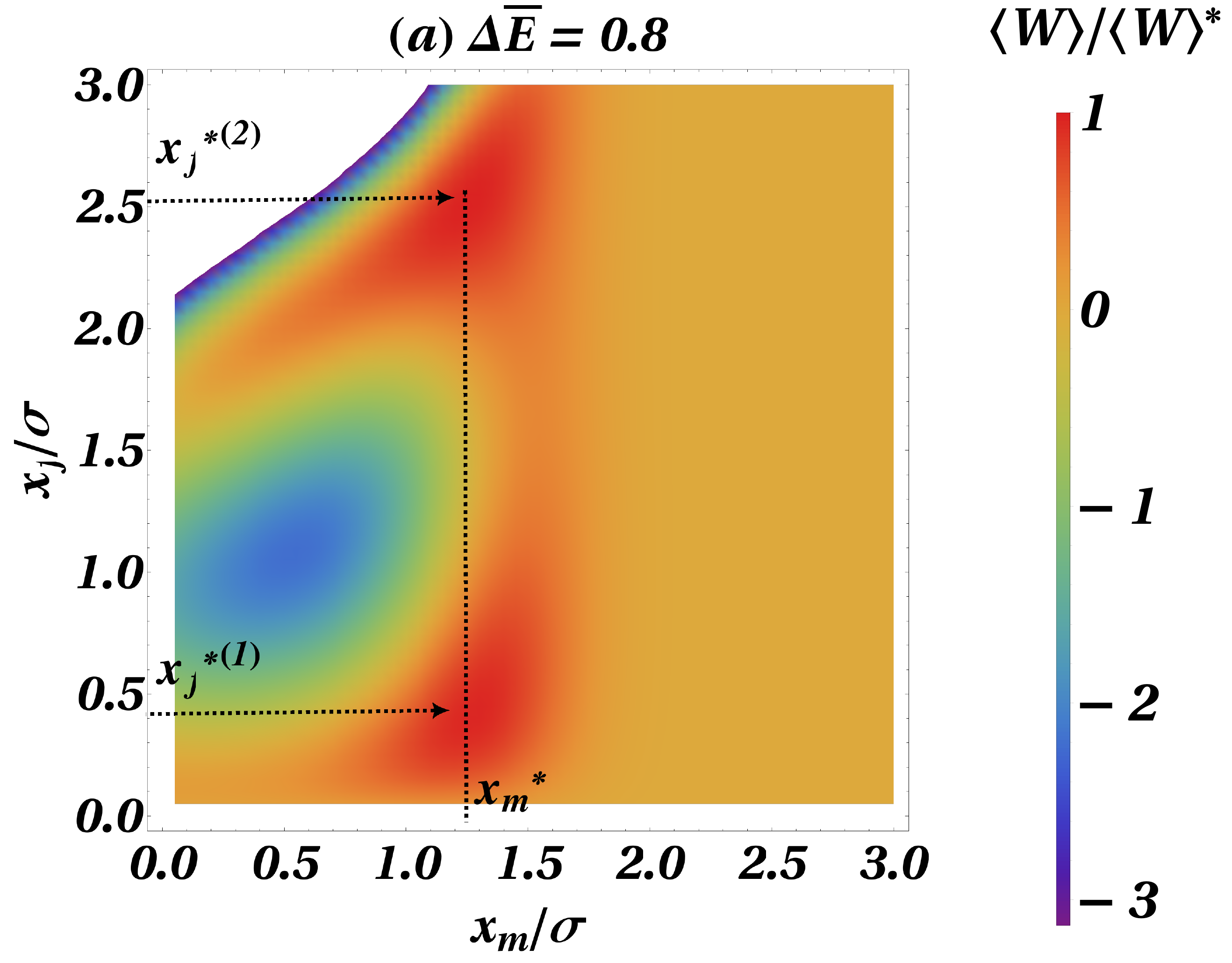}\\
\vspace{0.5cm}
\includegraphics[width=0.43\textwidth]{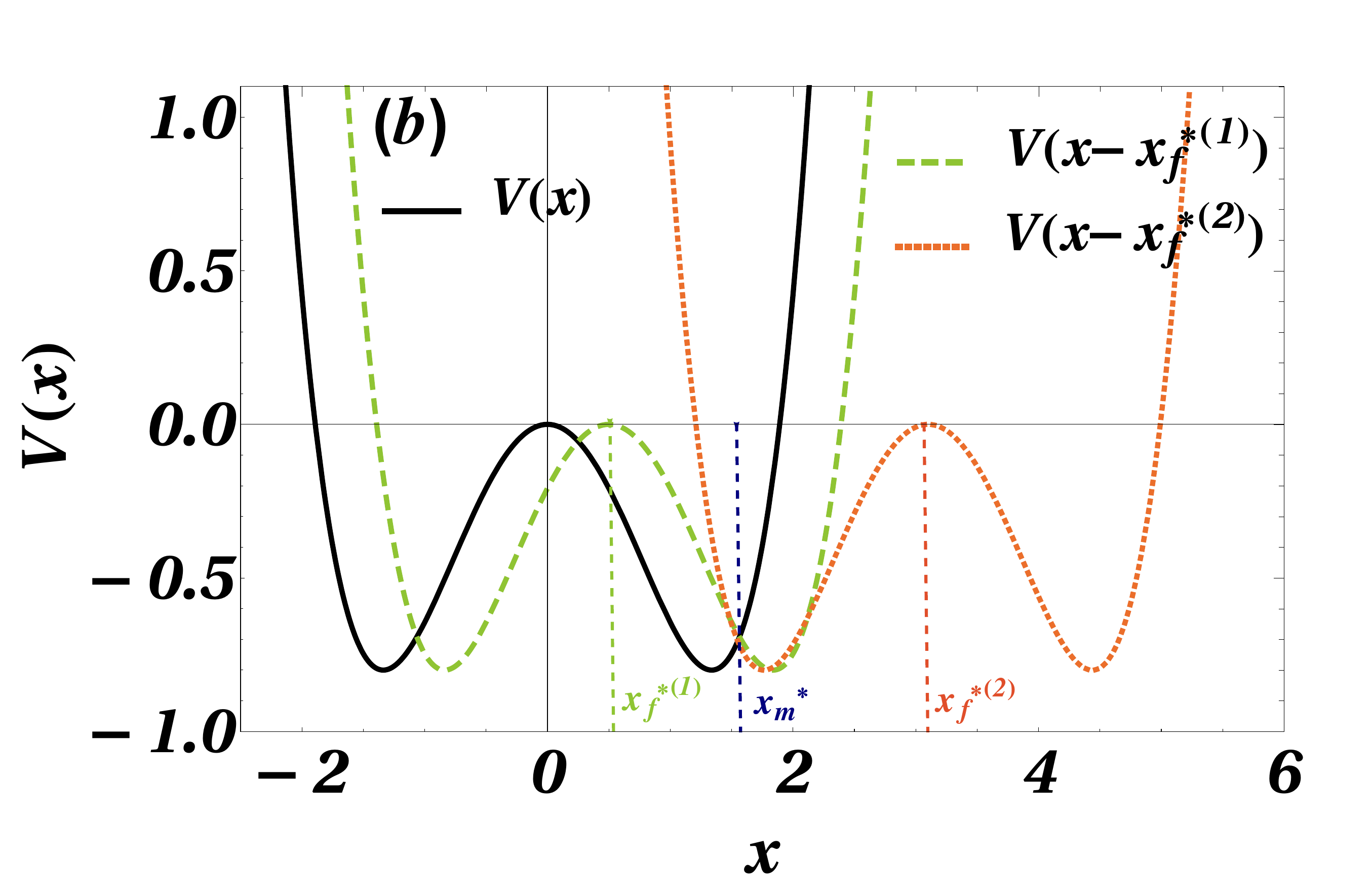}
\caption{(a) The density plot of the scaled work output $\left(\langle W \rangle / \langle W \rangle^* \right)$ under different scaled feedback control parameters, $x_m$ and $x_f$ for double-well potential of form $(V(x)= -\frac{a}{2}x^2+\frac{b}{4}x^4)$ with $\Delta \overline{E} = 0.8$. 
(b) Indication of related potential shift associated with the two distinct feedback locations $(x_{f}^{*(1)},x_{f}^{{*(2)}})$. Parameter set chosen: $b=1$, $\Delta \overline{E} = 0.8$, $\gamma = 1$ and $k_BT=1$ for all cases.}
\label{f6}
\end{figure}
We tune the energy barrier of a symmetric double well, $ \Delta E = \frac{a^2}{4b}$, and aim to evaluate the work extraction associated with the transition in confinement shape. Following the definition of equilibrium probability distribution $(P_{eq}(x))$ and standard deviation $(\sigma)$ (using Eq.~\ref{5} and \ref{9}), one can derive:
\begin{equation}\label{11}
\centering
    \begin{aligned}
      &P_{eq}(x) = \mathcal{N}_b  \exp \bigg [ \frac{\overline{a}}{2}x^2 - \frac{\overline{b}}{4} x^4 \bigg ], \\
       \text{and } 
      \sigma^2 &= \frac{I_{\frac{1}{4}}(\epsilon ) + 
    2 \epsilon  \left(I_{\frac{1}{4}}(\epsilon )+I_{\frac{3}{4}}(\epsilon )+I_{\frac{5}{4}}(\epsilon )+I_{-\frac{1}{4}}(\epsilon )\right)}{\sqrt{2} \sqrt{b \epsilon } \left(I_{\frac{1}{4}}(\epsilon )+I_{-\frac{1}{4}}(\epsilon )\right)}.
    \end{aligned}
\end{equation}
 Here, the normalisation constant $\mathcal{N}_b = \frac{e^{-\epsilon}}{\pi} \bigg ( \frac{2\overline{b}}{\epsilon} \bigg )^{\frac{1}{4}} \big [ I_{\frac{1}{4}} (\epsilon) +I_{-\frac{1}{4}} (\epsilon) \big ]$, $\epsilon = \Delta \overline{E}/2$ and $I_{\nu}(z)$ is the modified Bessel function of the first kind which takes the form, $I_{\nu}(z) = \sum_{k=0}^{\infty} \frac{(z/2)^{2k+\nu}}{\Gamma [k+ \nu + 1] k!}$. The $P_{eq}(x)$ shows a gradual change from unimodal to bimodal distribution as the energy barrier $(\Delta \overline{E} \ge 0)$ grows (Fig.~\ref{f3}(b)). In the limit of $\Delta \overline{E} \to 0$, $P_{eq}(x)$ exhibits a single peak distribution, signifying the sole contribution of the quartic term, i.e., $ V(x) = \frac{b}{4} x^4 $. With the increase in $ \Delta \overline{E}$ $(a > 0)$, $P_{eq}(x)$ evolves into a bimodal distribution, displaying two symmetric peaks separated by a well-defined minimum (at $x=0$). The separation between the two symmetric peaks grows with the scaled energy barrier, leading to a more dispersed system (Fig.~\ref{f3}(b)). The standard deviation of the equilibrium probability distribution, thus, increases with the rise in the scaled energy barrier (variation not shown here).
\paragraph*{}
  To examine the effect of such shape change of $P_{eq}(x)$ due to the modulation of the potential field on the optimal performance criteria and functionality of the information engine, we evaluate the average extractable work under varying $x_m$ and $x_f$. Fig.~\ref{f6}(a) presents a density plot illustrating the scaled work output $\left(-\langle W \rangle/\langle W \rangle^*\right)$ from a BIE with double-well confinement in a two-dimensional variation of scaled measurement distance $(x_m/ \sigma)$ and scaled feedback site $(x_f / \sigma)$ with $\Delta \overline{E} = 0.8$. $\langle W \rangle^*$ is the maximum extractable work under the given parameter choice. Interestingly, the density plot reveals two distinct regions of high work output, implying more than one set of optimal conditions for best performance. Two different sets of the optimal control parameters for the extraction of work are obtained in: (a) $x_m^* \sim 1.3 \sigma$ and $x_f^* \sim 2x_m^*$, and (b) $x_m^* \sim 1.3 \sigma$ and $x_f^* \sim 0.5x_m^*$. We identify that such multiple optimal feedback requisites arise from similar, though not identical, average potential changes during the feedback process. Fig.~\ref{f6}(b) presents a situation to depict the potential change for two good feedback sites related to the same optimal measurement distance $x_m^*$. One can readily notice that shifting the centre to two distinct feedback sites  $x_{f}^{*(1)}$ and $x_{f}^{*(2)}$ result in comparable average potential changes. Therefore, it is expected that information loss during the relaxation process $(\langle I_u\rangle)$ would be similar for such comparable potential environments because of the feedback. Fig.~\ref{6}(a) reveals another fascinating scenario in which a continuous increase in the feedback location induces the engine to transition from the heater to the refrigerator and back to the heater. We can notice such a re-entrance phenomenon for a suitable measurement distance (For example, $x_m/\sigma \sim 1$). 
\begin{figure}[!htb]
\centering
\includegraphics[width=0.43\textwidth]{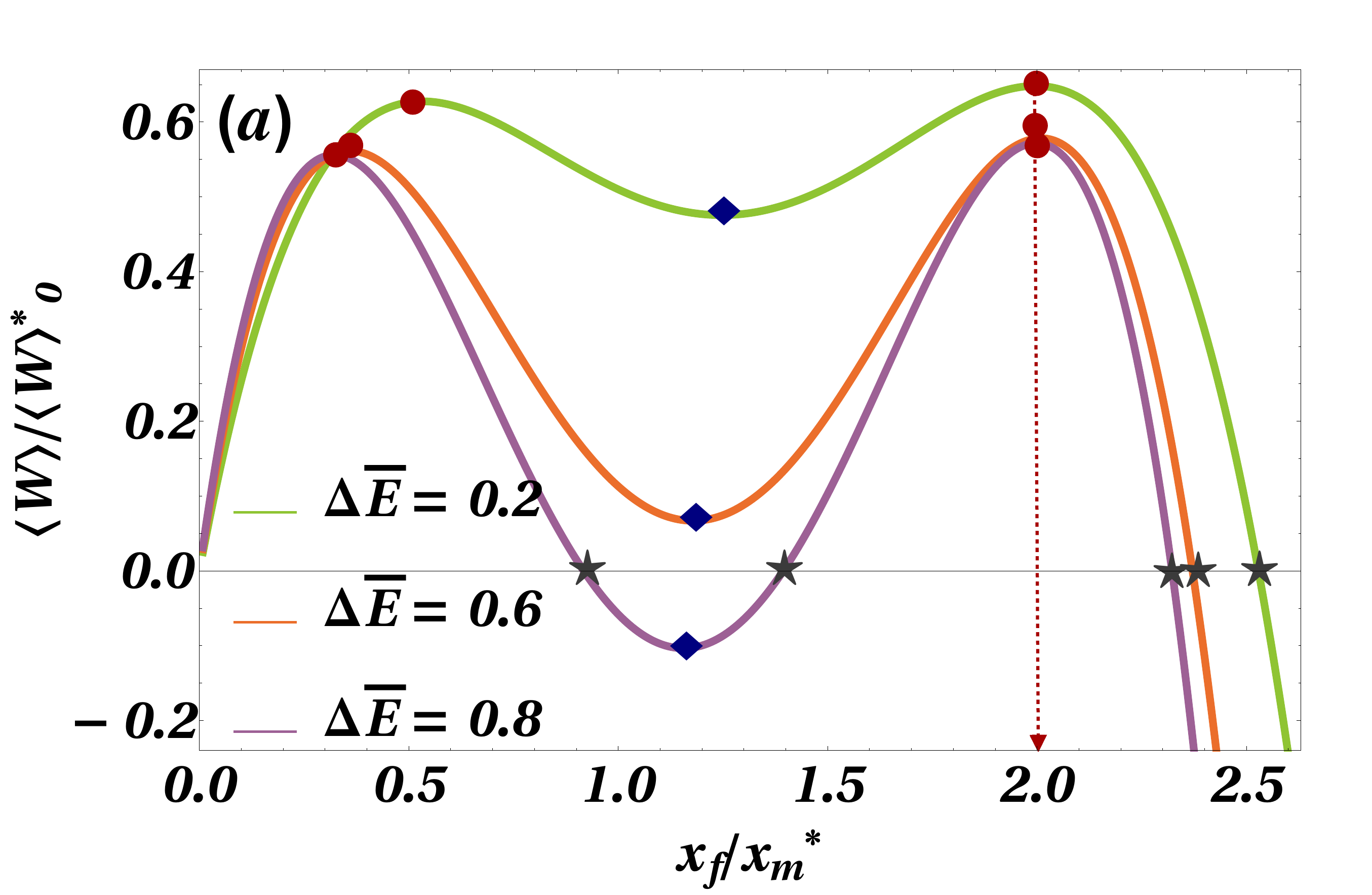}\\
\vspace{0.5cm}
\includegraphics[width=0.43\textwidth]{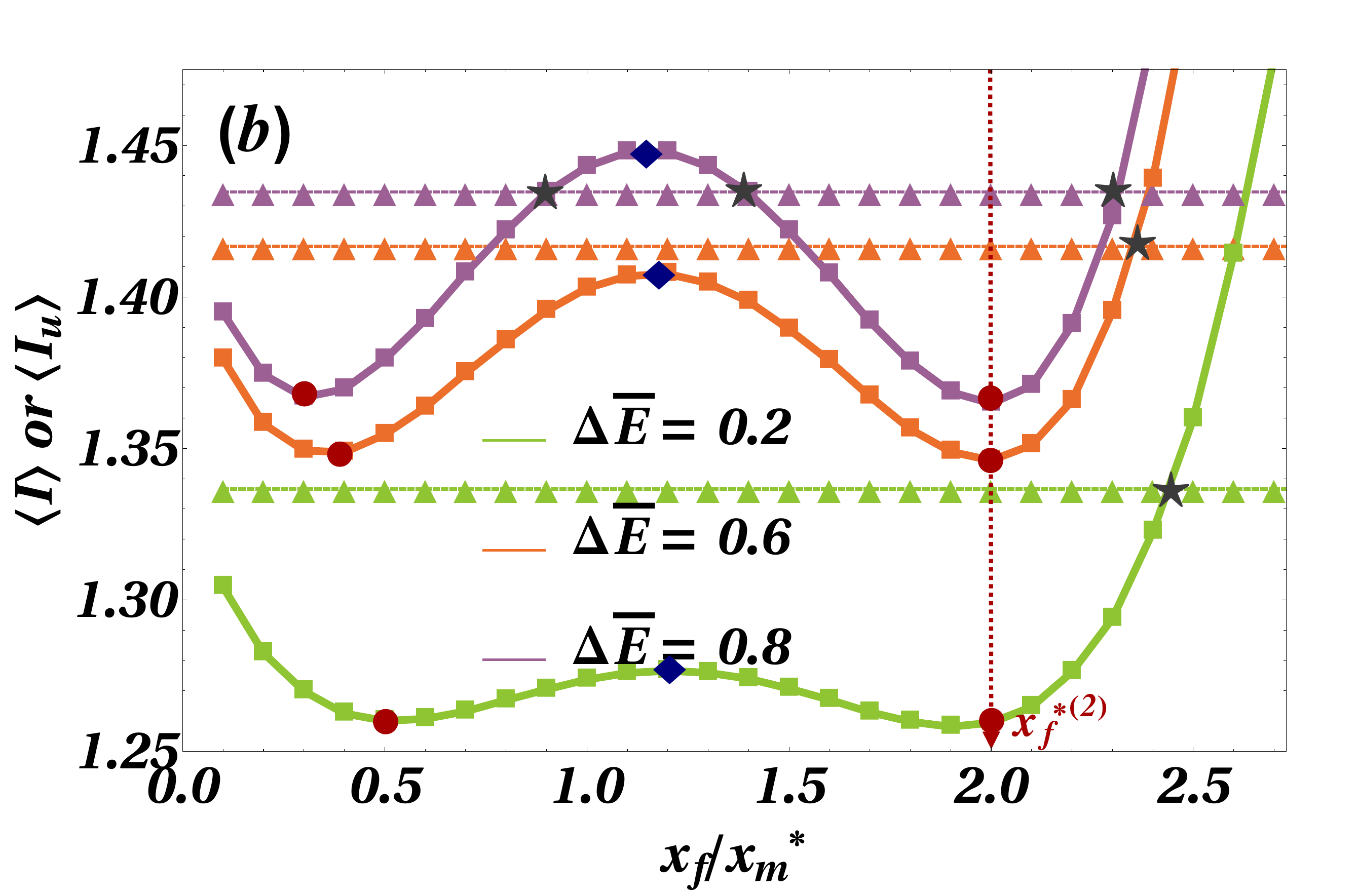}
\caption{(a)  Variation of a scaled work output  $\langle W \rangle / \langle W \rangle^*_0$, (b) average acquired information $(\langle I \rangle \text{, filled triangle-dotted lines })$ and average unavailable information $(\langle I_u \rangle \text{ filled square-solid lines })$, as a function of scaled feedback location $(x_f/x_m^*)$ with a given measurement length $x_m=x_{m}^{*}$ and for different $\Delta \overline{E}$. Red-coloured circles indicate feedback conditions to obtain output maxima, dark grey-coloured stars represent the output inversion points $(x_f^{inv})$ and blue-coloured diamonds indicate feedback conditions to obtain minimum work output. Parameter set is chosen: $b=1$, $\gamma =1$, and $k_BT=1$.}
\label{f7}
\end{figure}
\paragraph*{}
 Next, we plot the variation of scaled work $\langle W \rangle / \langle W \rangle^*_0$ as a function of the scaled feedback location $( x_f/x_m^*)$ across different potential landscapes with corresponding optimal measurement distance ($x_m=x_m^*$), as shown in Fig.~\ref{f7}(a). $\langle W \rangle_0^*$ refers to an optimal work obtained for $\Delta \overline{E} = 0$.  As expected from the previous density plot, the variation of scaled work $(\langle W \rangle / \langle W \rangle^*_0)$ with feedback location $( x_f/x_m^*)$ shows a non-monotonic trend characterized by two maxima and a minimum. We have two best (good) operating conditions at $x_f^* = 2x_m^*$ and at $x_f^* \sim 0.5 x_m^*$, respectively. We recall that the best measurement distance is unique $ x_m^* = C_m \sigma$. However, the proportionality constant $C_m$ and the dispersion $\sigma$ changes with increasing $\Delta \overline{E}$. Fig.~\ref{f7}(a) also shows that the introduction of a perturbed (unstable) potential centre reduces the amount of maximum work extraction $(\langle W \rangle^*)$. The two best operating feedback sites are separated by a feedback site that produces an output minimum. The depth of the minimum increases with increasing instability at the potential centre. In the limit of $(\Delta \overline{E} \to \text{high})$, BIE exhibits heater-to-refrigeration and is followed by a re-entrance for an intermediate feedback location. When the feedback site is too long, we obtain a heater-to-refrigerator transition as expected.
\paragraph*{}To explain the observed performance reentry and other disparity in energy harvesting under the different extent of bi-stability in the confinement, as observed in Figs.~\ref{f6}(a) and \ref{f7}(a),  we calculate the average acquired information $(\langle I \rangle)$ and the average unavailable information $(\langle I_u \rangle)$ following Eqs.~\ref{6}, \ref{7} and \ref{11}. Fig.~\ref{f7}(b) shows the variation of $\langle I \rangle$ and $\langle I_u \rangle$ as a function of the scaled feedback location $( x_f/x_m^*)$ for the different extent of bistability in confinement and at the best measurement distance $(x_m=x_m^*)$. As is obvious from the definition, for a given protocol control and for an unchanged barrier height, $\langle I \rangle$ is invariant to the location of the feedback. On the other hand, $ \langle I_u \rangle $ shows a non-monotonic variation with increasing $x_f$. A comparison between Fig.~\ref{f7}(a) and Fig.~\ref{f7}(b) shows that the loss of information during relaxation is minimized under feedback conditions, maximizing the extraction of work. The amount of information available $(\langle I \rangle - \langle I_u \rangle)$ in the best feedback location decreases with increasing instability of the potential centre. For certain feedback sites, $\langle I_u \rangle > \langle I \rangle$, that leads to an inversion in the sign of output work, and the engine runs as a refrigerator. Therefore, the instability in the potential centre leads to a non-trivial variation of the information loss during the relaxation. For a given height of such centrosymmetric potential hills, a careful variation of the feedback locations may result in multiple situations of best work extraction, and a BIE may undergo a heater-refrigerator and hence a re-entrance event with increasing $x_f$.
 
\subsection{Modulation of a monostable trap to a triple well confinement}
\begin{figure}[!htb]
    \centering    
\includegraphics[width=0.44\textwidth]{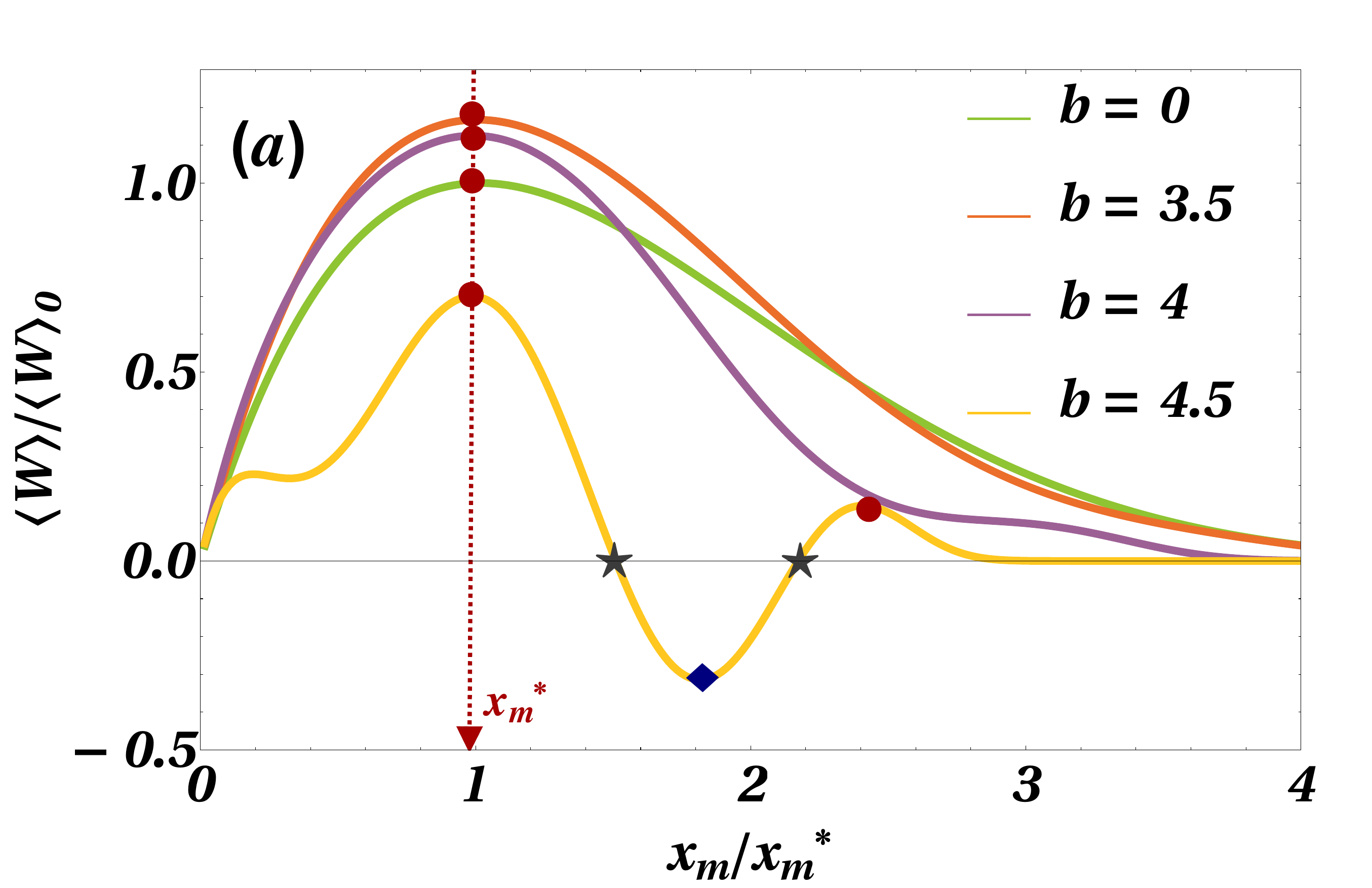}\\
\vspace{0.5cm}
\includegraphics[width=0.44\textwidth]{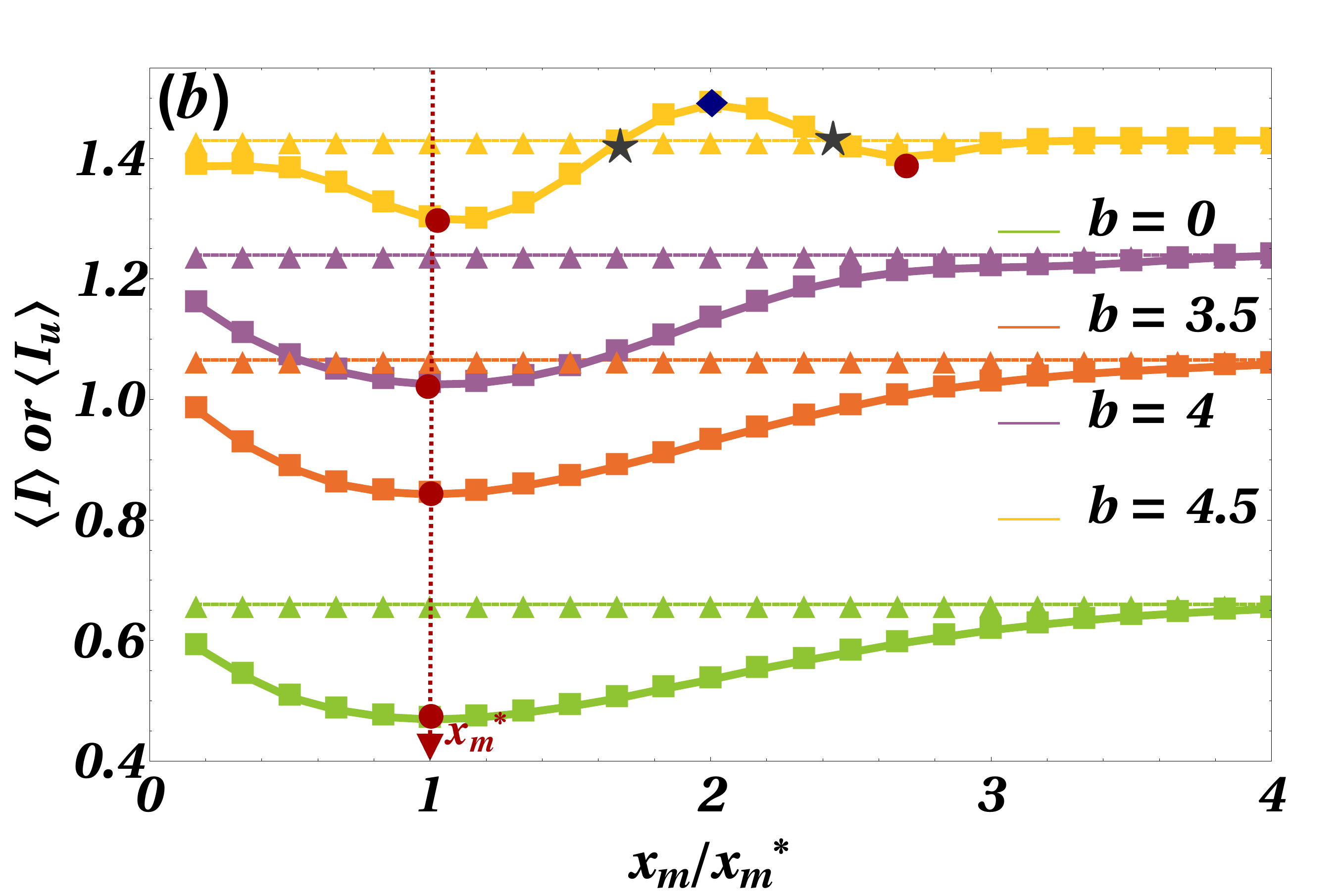}
    \caption{(a)  Variation of a scaled work output $(\langle W \rangle/\langle W \rangle_0)$, (b) average acquired information $(\langle I \rangle \text{, filled triangle-dotted lines })$ and average unavailable information $(\langle I_u \rangle \text{, filled square-solid lines })$, as a function of scaled measurement length $(x_m/x_m^*)$ with $x_{f} = 2x_{m}$, for different values of $b$ of the potential of form $V(x)= \frac{a}{2}x^2-\frac{b}{4}x^4 + \frac{c}{6}x^6$. The red coloured circle indicates the location of a maximum work $-\langle W \rangle^*$, the dark grey coloured stars represent inversion points and blue coloured diamonds indicate a minimum refrigeration work. Parameter set chosen: $a=4$, $c=1$, $\gamma =1$ and $k_BT=1$.}   
    \label{f9}
\end{figure}

Finally, to understand the information-energy exchange in a BIE that operates in multi-stable confinement, we tune the shape of the potential from monostable to tristable potential. One of the convenient approaches involves modifying the quartic contribution of the potential profile of the form: $V(x) = \frac{ax^2}{2}-\frac{b x^4}{4}+\frac{c x^6}{6}$, where $a$, $b$ and $c$ are constant (positive) potential parameters. In the limit of $b \to 0$ the potential is mono-stable. When the nonzero quartic contribution approaches the saddle point condition $b \sim \sqrt{4ac}$, the potential shows a monostability at the centre with two centrosymmetric concave shoulders. As the quartic contribution increases further, $b > \sqrt{4ac}$, the potential transitions into a tristable form with a stable centre and two symmetrically placed wells. 
The corresponding equilibrium probability distribution $P_{eq}(x)$ (using Eq.~\ref{5})  shows a transition from a unimodal to a trimodal steady-state distribution as the contribution of the quartic term varies (shown in Fig.~\ref{3}(c)). Clearly, the standard deviation of the probability distribution $(\sigma)$ increases with an increase in the quartic contribution $(b)$.
\begin{figure}[!htb]
    \centering    
\includegraphics[width=0.45\textwidth]{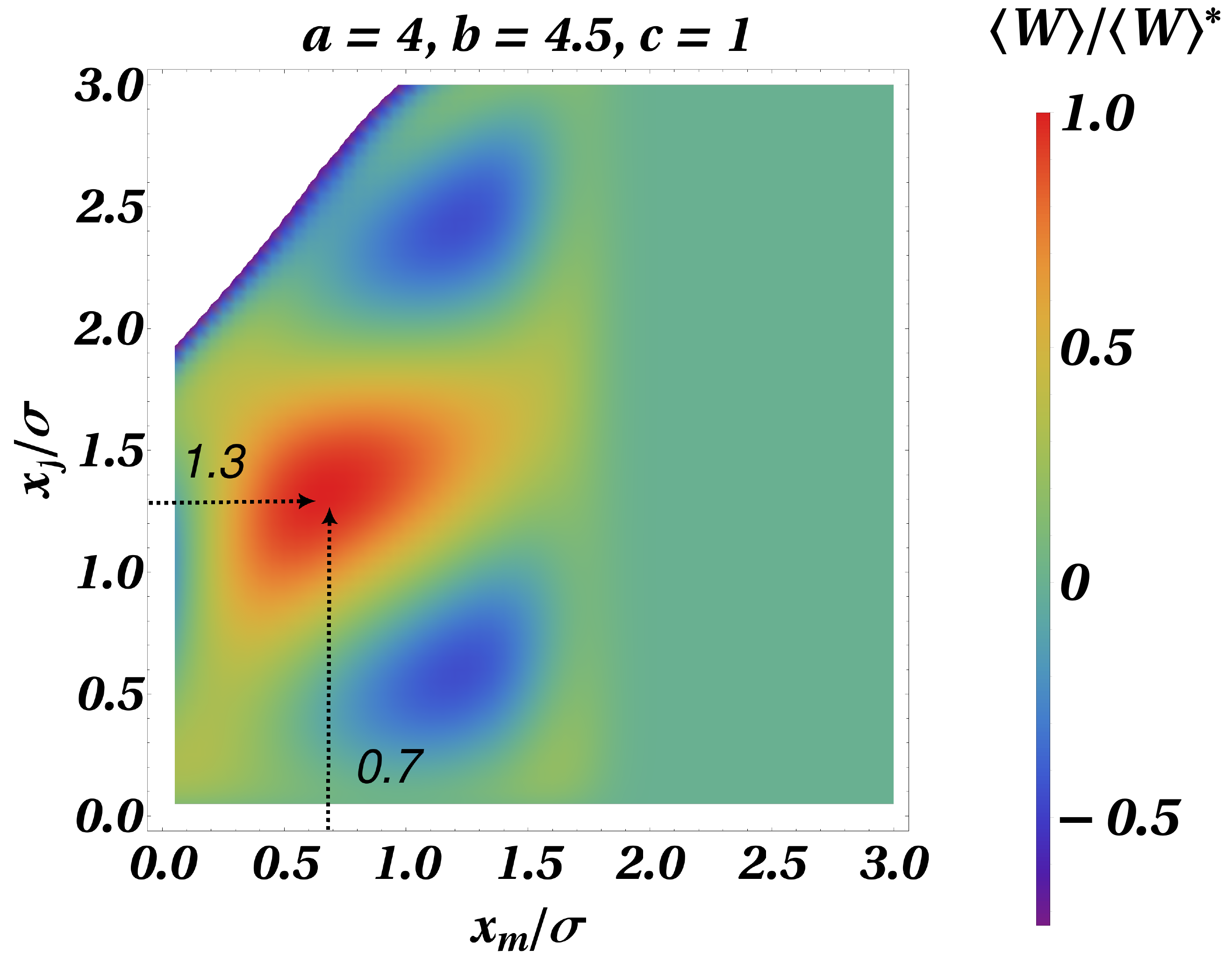}
    \caption{The density plot of scaled work output $(\langle W \rangle / \langle W \rangle^*)$ under different scaled feedback control parameters, $x_m/\sigma$ and $x_f/\sigma$ for triple-well potential of form $V(x) = \frac{ax^2}{2}-\frac{b x^4}{4}+\frac{c x^6}{6}$ with $\gamma =1$ and $k_BT=1$.}   
    \label{f8}
\end{figure}
To assess the optimal performance criteria and functionality of the information engine with such multiple potential, we obtained the density plot of the scaled work output $(\langle W \rangle/\langle W \rangle^*)$ by systematically varying measurement distance $(x_m/ \sigma)$ and feedback site $(x_f / \sigma)$, see Fig.~\ref{f8}. For a given confinement parameters $(a=4, \; b=4.5, \;c=1)$, the optimal requisites for maximum work extraction read as  $x_m^* \sim 0.7 \sigma$ and $x_f^* = 2x_m^*$. 
Interestingly, with a suitable nonzero measurement distance $(x_m/\sigma \sim 1)$, increasing the feedback location causes the engine to transition from a refrigerator to a heater and revert to a refrigerator, demonstrating a re-entrance behaviour. A comparable examination of the BIE with different potential parameters reveals the best conditions for work extraction always remain at $x_m^* = C_m \sigma$ and $ x_f^* = 2x_m^*$.
\paragraph*{}
Next, we focus on how the maximum work extraction $ -\langle W \rangle^*$ (with a feedback site at $x_f=2x_m$) depends on the measurement distance and the quartic contribution ($b$) of the multistable trapping. Fig.~\ref{f9}(a) presents the scaled work $( -\langle W \rangle / \langle W \rangle_0 )$ versus the scaled measurement distance $x_m/x_m^*$ for potential with different values of $b$, where $\langle W \rangle_0$ is the work output at $b = 0$.
The variation of scaled work $\langle W \rangle / \langle W \rangle_0 $ with the measurement distance exhibits a non-monotonic trend for all choices of $b$. For any arbitrary choice of $b$, a single set of optimal conditions for maximum work extraction $( \langle W \rangle^*)$, given as: $x_m^* = C_m \sigma$ and $x_f^* = 2x_m^*$. If $b$ is not  too high, we find that maximum work extraction $\langle \overline{W} \rangle^*$ increases with the quartic contribution of the sextic confinement $(b>0)$. However, a further increase in the quartic contribution $(b > \sqrt{4ac})$ leads to a decline in $\langle W \rangle^*$. A potential confinement with two concave shoulders gives higher $\langle W \rangle^*$. In the limit of a BIE with pronounced triple-wells, a heater-to-refrigerator re-entrance occurs with increasing measurement distance.

To explore the nontrivial work output for different tristable potential profiles, we examine the total information $\langle I \rangle$ and the unavailable information $\langle I_u \rangle$  using their definitions as found in Eq.~\ref{6} and \ref{7}. Fig.~\ref{f9}(b) depicts the variation of $\langle I \rangle$ and $\langle I_u \rangle$ as a function of scaled measurement distance $(x_m/x_m^*)$ keeping feedback location as $x_f=2x_m$. The average acquired information during the measurement step remains invariant to the measurement distance $x_m$, as expected. However, the information lost during the relaxation step follows a non-monotonic but non-trivial trend with increasing measurement distance. A one to one comparison between Fig.~\ref{f9}(a) and Fig.~\ref{f9}(b) reveals that the amount of information lost during the relaxation governs the extent of output work.
The maximum work extraction setup is, thus, always associated with the least information loss (minimum $\langle I_u \rangle$), irrespective of the choice of $b$. Fig.~\ref{9} also indicates that the engine functions as a refrigerator whenever the information lost during the relaxation step dominates the total information acquired during the measurement $(\langle I \rangle < \langle I_u \rangle)$. Therefore, a proper tuning on the potential may lead to a heater transition re-entrance phenomena with varying the measurement distance. For the present variation, tristability with $a=4, \; b=4.5, \;c=1$ depicts one such scenario. Finally, when the multistable confinement is dominated by the monostable trapping around the centre (low $b$), the loss in $\langle I_u \rangle$ at its minimum when two concave shoulders are formed. The creation of two such concave potential shoulders leads to such an enhancement of maximum output value. Qualitatively, the enhancement continues until a saddle point situation ($b\sim \sqrt{4ac}$).

\section{SUMMARY}
We examine the effect of shape modulation of centrosymmetric confining potentials of Brownian information engines operating under an asymmetric feedback cycle. The output work and its optimal requirements in terms of the reference measurement distance $(x_m)$ and the feedback site $(x_f)$ are greatly influenced by the change in shapes of the potential trap. The BIE with monostable confinement of form $V(x)=a|x|^n,(n > 0)$ exhibits optimal condition for maximum work at a measurement distance $x_m^* = C_m \sigma$ and a feedback location at $x_f^*=2 x_m^*$ for all $n$. Overall, the loss of information during the relaxation process decreases with increased concavity of potential $(n < 1)$ that enhances information-energy exchange.
The engine shows a transition from heater to refrigeration beyond specific values of feedback location ($x_f > x_f^{inv}$).
Both the standard deviation $\sigma$ and the inversion feedback distance $x_f^{inv}$ increase, while the proportionality constant $C_m$  decreases non-linearly with an increase in concavity (decrease in the power exponent $n$). 
\paragraph*{}Next, we analyze the effect of an unstable potential centre in confining potential on information-energy exchange and consequent work harvesting under the same asymmetric feedback. Because of the centrosymmetric nature of the trapping, one may find multiple (two for a bistable potential) feedback sites with similar (not exact) relaxation environments.
Consequently, two distinct sets of optimal conditions could be obtained for maximum extraction of work. With a given (optimized) measurement distance, the output work shows an interesting nonmonotonic trend with increasing feedback location ($x_f$), characterized by two maxima and a minimum.  The maximum work extraction decreases with the increasing scaled energy barrier of the bistable confinement, which can be attributed to a sudden gain in potential during the feedback process due to the unstability of the confinement centre. With a careful choice of control parameters, the feedback location that results in a minimum work output may cause a refrigeration effect. Therefore,  double-well potential with deeper wells $(\Delta \overline{E} \to \text{high})$ exhibits a fascinating heater-to-refrigeration transition and followed by a re-entrance event upon changing the $x_f$. Beyond a high value of the feedback distance, $x_f \to \text{high}$, the BIE always acts as a refrigerator, as expected. 

Finally, we investigated a continuous switching from monostable to tristable potential to comprehend the implication of potential with multiple energy valleys (stable) and unstable hills on engine functionality. The amount of extracted work at the optimum measurement distance shows a turn-over, passing through a maximum value during this potential switch. The potential landscape with concave shoulders will have reduced information loss during relaxation, yielding higher work extraction under optimal conditions. For confinement with pronounced basins, BIE shows heater-to-refrigeration re-entrance phenomena with varying measurement distances. We believe that the present study opens up the scope of further experimental and theoretical investigations where suitable tuning on confined potential shapes in engineering the best performance and controlling the functionality of Brownian information engines is needed. 
\begin{acknowledgments}
RR acknowledges IIT Tirupati for fellowship [DST/INSPIRE/03/2021/002138]. DM thanks SERB (Project No. ECR/2018/002830/CS), Department of Science and Technology, Government of India, for financial support and IIT Tirupati for the computational facility.
\end{acknowledgments}
\section*{Data Availability}
The data that support the findings of this study are available within the article.
\bibliographystyle{achemso}
\bibliography{References}

\providecommand{\latin}[1]{#1}
\makeatletter
\providecommand{\doi}
  {\begingroup\let\do\@makeother\dospecials
  \catcode`\{=1 \catcode`\}=2 \doi@aux}
\providecommand{\doi@aux}[1]{\endgroup\texttt{#1}}
\makeatother
\providecommand*\mcitethebibliography{\thebibliography}
\csname @ifundefined\endcsname{endmcitethebibliography}  {\let\endmcitethebibliography\endthebibliography}{}
\begin{mcitethebibliography}{92}
\providecommand*\natexlab[1]{#1}
\providecommand*\mciteSetBstSublistMode[1]{}
\providecommand*\mciteSetBstMaxWidthForm[2]{}
\providecommand*\mciteBstWouldAddEndPuncttrue
  {\def\EndOfBibitem{\unskip.}}
\providecommand*\mciteBstWouldAddEndPunctfalse
  {\let\EndOfBibitem\relax}
\providecommand*\mciteSetBstMidEndSepPunct[3]{}
\providecommand*\mciteSetBstSublistLabelBeginEnd[3]{}
\providecommand*\EndOfBibitem{}
\mciteSetBstSublistMode{f}
\mciteSetBstMaxWidthForm{subitem}{(\alph{mcitesubitemcount})}
\mciteSetBstSublistLabelBeginEnd
  {\mcitemaxwidthsubitemform\space}
  {\relax}
  {\relax}

\bibitem[Rex(2017)]{rex2017maxwell}
Rex,~A. Maxwell’s demon — A historical review. \emph{Entropy} \textbf{2017}, \emph{19}, 240\relax
\mciteBstWouldAddEndPuncttrue
\mciteSetBstMidEndSepPunct{\mcitedefaultmidpunct}
{\mcitedefaultendpunct}{\mcitedefaultseppunct}\relax
\EndOfBibitem
\bibitem[Maxwell and Pesic(2001)Maxwell, and Pesic]{maxwell2001theory}
Maxwell,~J.~C.; Pesic,~P. \emph{Theory of heat}; Courier Corporation, 2001\relax
\mciteBstWouldAddEndPuncttrue
\mciteSetBstMidEndSepPunct{\mcitedefaultmidpunct}
{\mcitedefaultendpunct}{\mcitedefaultseppunct}\relax
\EndOfBibitem
\bibitem[Szilard(1929)]{Szilard1929zphys}
Szilard,~L. {\"u}ber die {E}ntropieverminderung in einem thermodynamischen {S}ystem bei {E}ingriffen intelligenter Wesen. \emph{Z. Phys} \textbf{1929}, \emph{53}, 840\relax
\mciteBstWouldAddEndPuncttrue
\mciteSetBstMidEndSepPunct{\mcitedefaultmidpunct}
{\mcitedefaultendpunct}{\mcitedefaultseppunct}\relax
\EndOfBibitem
\bibitem[Landauer(1961)]{Landaueribm1961}
Landauer,~R. Irreversibility and {H}eat {G}eneration in the {C}omputing {P}rocess. \emph{IBM J. Res. Dev} \textbf{1961}, \emph{5}, 183\relax
\mciteBstWouldAddEndPuncttrue
\mciteSetBstMidEndSepPunct{\mcitedefaultmidpunct}
{\mcitedefaultendpunct}{\mcitedefaultseppunct}\relax
\EndOfBibitem
\bibitem[Seifert(2005)]{seifert2005}
Seifert,~U. Entropy Production along a Stochastic Trajectory and an Integral Fluctuation Theorem. \emph{Phys. Rev. Lett.} \textbf{2005}, \emph{95}, 040602\relax
\mciteBstWouldAddEndPuncttrue
\mciteSetBstMidEndSepPunct{\mcitedefaultmidpunct}
{\mcitedefaultendpunct}{\mcitedefaultseppunct}\relax
\EndOfBibitem
\bibitem[Seifert(2008)]{seifert2008}
Seifert,~U. Stochastic thermodynamics: principles and perspectives. \emph{Euro. Phys. J. B} \textbf{2008}, \emph{64}, 423\relax
\mciteBstWouldAddEndPuncttrue
\mciteSetBstMidEndSepPunct{\mcitedefaultmidpunct}
{\mcitedefaultendpunct}{\mcitedefaultseppunct}\relax
\EndOfBibitem
\bibitem[Sagawa and Ueda(2008)Sagawa, and Ueda]{Sagwa2008prl}
Sagawa,~T.; Ueda,~M. Second Law of Thermodynamics with Discrete Quantum Feedback Control. \emph{Phys. Rev. Lett.} \textbf{2008}, \emph{100}, 080403\relax
\mciteBstWouldAddEndPuncttrue
\mciteSetBstMidEndSepPunct{\mcitedefaultmidpunct}
{\mcitedefaultendpunct}{\mcitedefaultseppunct}\relax
\EndOfBibitem
\bibitem[Sagawa and Ueda(2010)Sagawa, and Ueda]{Sagawa2010prl}
Sagawa,~T.; Ueda,~M. Generalized {J}arzynski {E}quality under {N}onequilibrium {F}eedback {C}ontrol. \emph{Phys. Rev. Lett.} \textbf{2010}, \emph{104}, 090602\relax
\mciteBstWouldAddEndPuncttrue
\mciteSetBstMidEndSepPunct{\mcitedefaultmidpunct}
{\mcitedefaultendpunct}{\mcitedefaultseppunct}\relax
\EndOfBibitem
\bibitem[Sagawa and Ueda(2009)Sagawa, and Ueda]{Sagawa2009prl}
Sagawa,~T.; Ueda,~M. Minimal {E}nergy {C}ost for {T}hermodynamic {I}nformation {P}rocessing: {M}easurement and {I}nformation {E}rasure. \emph{Phys. Rev. Lett.} \textbf{2009}, \emph{102}, 250602\relax
\mciteBstWouldAddEndPuncttrue
\mciteSetBstMidEndSepPunct{\mcitedefaultmidpunct}
{\mcitedefaultendpunct}{\mcitedefaultseppunct}\relax
\EndOfBibitem
\bibitem[Sagawa and Ueda(2012)Sagawa, and Ueda]{Sagawa2012pre}
Sagawa,~T.; Ueda,~M. Nonequilibrium thermodynamics of feedback control. \emph{Phys. Rev. E} \textbf{2012}, \emph{85}, 021104\relax
\mciteBstWouldAddEndPuncttrue
\mciteSetBstMidEndSepPunct{\mcitedefaultmidpunct}
{\mcitedefaultendpunct}{\mcitedefaultseppunct}\relax
\EndOfBibitem
\bibitem[Horowitz and Vaikuntanathan(2010)Horowitz, and Vaikuntanathan]{Horowitz2010pre}
Horowitz,~J.~M.; Vaikuntanathan,~S. Nonequilibrium detailed fluctuation theorem for repeated discrete feedback. \emph{Phys. Rev. E} \textbf{2010}, \emph{82}, 061120\relax
\mciteBstWouldAddEndPuncttrue
\mciteSetBstMidEndSepPunct{\mcitedefaultmidpunct}
{\mcitedefaultendpunct}{\mcitedefaultseppunct}\relax
\EndOfBibitem
\bibitem[Horowitz and Parrondo(2011)Horowitz, and Parrondo]{Horowitz2011epl}
Horowitz,~J.~M.; Parrondo,~J.~M. Thermodynamic reversibility in feedback processes. \emph{Europhys. Lett.} \textbf{2011}, \emph{95}, 10005\relax
\mciteBstWouldAddEndPuncttrue
\mciteSetBstMidEndSepPunct{\mcitedefaultmidpunct}
{\mcitedefaultendpunct}{\mcitedefaultseppunct}\relax
\EndOfBibitem
\bibitem[Jarzynski(1997)]{Jarzynski1997prl}
Jarzynski,~C. Nonequilibrium Equality for Free Energy Differences. \emph{Phys. Rev. Lett.} \textbf{1997}, \emph{78}, 2690\relax
\mciteBstWouldAddEndPuncttrue
\mciteSetBstMidEndSepPunct{\mcitedefaultmidpunct}
{\mcitedefaultendpunct}{\mcitedefaultseppunct}\relax
\EndOfBibitem
\bibitem[Jarzynski(2011)]{jarzynski2011}
Jarzynski,~C. Equalities and Inequalities: Irreversibility and the Second Law of Thermodynamics at the Nanoscale. \emph{Annu. Rev. Condens. Matter Phys.} \textbf{2011}, \emph{2}, 329--351\relax
\mciteBstWouldAddEndPuncttrue
\mciteSetBstMidEndSepPunct{\mcitedefaultmidpunct}
{\mcitedefaultendpunct}{\mcitedefaultseppunct}\relax
\EndOfBibitem
\bibitem[Seifert(2012)]{Seifert_2012}
Seifert,~U. Stochastic thermodynamics, fluctuation theorems and molecular machines. \emph{Rep. Prog. Phys.} \textbf{2012}, \emph{75}, 126001\relax
\mciteBstWouldAddEndPuncttrue
\mciteSetBstMidEndSepPunct{\mcitedefaultmidpunct}
{\mcitedefaultendpunct}{\mcitedefaultseppunct}\relax
\EndOfBibitem
\bibitem[Gaspard(2004)]{Gaspard2004}
Gaspard,~P. Fluctuation theorem for nonequilibrium reactions. \emph{J. Chem. Phys.} \textbf{2004}, \emph{120}, 8898\relax
\mciteBstWouldAddEndPuncttrue
\mciteSetBstMidEndSepPunct{\mcitedefaultmidpunct}
{\mcitedefaultendpunct}{\mcitedefaultseppunct}\relax
\EndOfBibitem
\bibitem[Das \latin{et~al.}(2012)Das, Mondal, and Ray]{das2012shape}
Das,~M.; Mondal,~D.; Ray,~D.~S. Shape change as entropic phase transition: A study using Jarzynski relation. \emph{J. Chem. Sci.} \textbf{2012}, \emph{124}, 21\relax
\mciteBstWouldAddEndPuncttrue
\mciteSetBstMidEndSepPunct{\mcitedefaultmidpunct}
{\mcitedefaultendpunct}{\mcitedefaultseppunct}\relax
\EndOfBibitem
\bibitem[Abreu and Seifert(2011)Abreu, and Seifert]{Abreu2011epl}
Abreu,~D.; Seifert,~U. Extracting work from a single heat bath through feedback. \emph{Europhys. Lett.} \textbf{2011}, \emph{94}, 10001\relax
\mciteBstWouldAddEndPuncttrue
\mciteSetBstMidEndSepPunct{\mcitedefaultmidpunct}
{\mcitedefaultendpunct}{\mcitedefaultseppunct}\relax
\EndOfBibitem
\bibitem[Abreu and Seifert(2012)Abreu, and Seifert]{Abreu2012prl}
Abreu,~D.; Seifert,~U. Thermodynamics of {G}enuine {N}onequilibrium {S}tates under {F}eedback {C}ontrol. \emph{Phys. Rev. Lett.} \textbf{2012}, \emph{108}, 030601\relax
\mciteBstWouldAddEndPuncttrue
\mciteSetBstMidEndSepPunct{\mcitedefaultmidpunct}
{\mcitedefaultendpunct}{\mcitedefaultseppunct}\relax
\EndOfBibitem
\bibitem[Bauer \latin{et~al.}(2012)Bauer, Abreu, and Seifert]{Bauer_2012}
Bauer,~M.; Abreu,~D.; Seifert,~U. Efficiency of a Brownian information machine. \emph{J. Phys. A Math. Theor.} \textbf{2012}, \emph{45}, 162001\relax
\mciteBstWouldAddEndPuncttrue
\mciteSetBstMidEndSepPunct{\mcitedefaultmidpunct}
{\mcitedefaultendpunct}{\mcitedefaultseppunct}\relax
\EndOfBibitem
\bibitem[Mandal \latin{et~al.}(2013)Mandal, Quan, and Jarzynski]{Mandal_2013}
Mandal,~D.; Quan,~H.~T.; Jarzynski,~C. Maxwell's Refrigerator: An Exactly Solvable Model. \emph{Phys. Rev. Lett.} \textbf{2013}, \emph{111}, 030602\relax
\mciteBstWouldAddEndPuncttrue
\mciteSetBstMidEndSepPunct{\mcitedefaultmidpunct}
{\mcitedefaultendpunct}{\mcitedefaultseppunct}\relax
\EndOfBibitem
\bibitem[Kosugi(2013)]{Taichi_2013}
Kosugi,~T. Perpetual extraction of work from a nonequilibrium dynamical system under Markovian feedback control. \emph{Phys. Rev. E} \textbf{2013}, \emph{88}, 032144\relax
\mciteBstWouldAddEndPuncttrue
\mciteSetBstMidEndSepPunct{\mcitedefaultmidpunct}
{\mcitedefaultendpunct}{\mcitedefaultseppunct}\relax
\EndOfBibitem
\bibitem[Pal \latin{et~al.}(2014)Pal, Rana, Saha, and Jayannavar]{Pal2014pre}
Pal,~P.~S.; Rana,~S.; Saha,~A.; Jayannavar,~A.~M. Extracting work from a single heat bath: A case study of a Brownian particle under an external magnetic field in the presence of information. \emph{Phys. Rev. E} \textbf{2014}, \emph{90}, 022143\relax
\mciteBstWouldAddEndPuncttrue
\mciteSetBstMidEndSepPunct{\mcitedefaultmidpunct}
{\mcitedefaultendpunct}{\mcitedefaultseppunct}\relax
\EndOfBibitem
\bibitem[Ashida \latin{et~al.}(2014)Ashida, Funo, Murashita, and Ueda]{Ashida2014pre}
Ashida,~Y.; Funo,~K.; Murashita,~Y.; Ueda,~M. General achievable bound of extractable work under feedback control. \emph{Phys. Rev. E} \textbf{2014}, \emph{90}, 052125\relax
\mciteBstWouldAddEndPuncttrue
\mciteSetBstMidEndSepPunct{\mcitedefaultmidpunct}
{\mcitedefaultendpunct}{\mcitedefaultseppunct}\relax
\EndOfBibitem
\bibitem[Ali \latin{et~al.}(2022)Ali, Rafeek, and Mondal]{ali2022geometric}
Ali,~S.~Y.; Rafeek,~R.; Mondal,~D. Geometric Brownian information engine: Upper bound of the achievable work under feedback control. \emph{J. Chem. Phys.} \textbf{2022}, \emph{156}, 014902\relax
\mciteBstWouldAddEndPuncttrue
\mciteSetBstMidEndSepPunct{\mcitedefaultmidpunct}
{\mcitedefaultendpunct}{\mcitedefaultseppunct}\relax
\EndOfBibitem
\bibitem[Rafeek \latin{et~al.}(2023)Rafeek, Ali, and Mondal]{rafeek2023geometric}
Rafeek,~R.; Ali,~S.~Y.; Mondal,~D. Geometric Brownian information engine: Essentials for the best performance. \emph{Phys. Rev. E} \textbf{2023}, \emph{107}, 044122\relax
\mciteBstWouldAddEndPuncttrue
\mciteSetBstMidEndSepPunct{\mcitedefaultmidpunct}
{\mcitedefaultendpunct}{\mcitedefaultseppunct}\relax
\EndOfBibitem
\bibitem[Park \latin{et~al.}(2016)Park, Lee, and Noh]{Park2016pre}
Park,~J.~M.; Lee,~J.~S.; Noh,~J.~D. Optimal tuning of a confined Brownian information engine. \emph{Phys. Rev. E} \textbf{2016}, \emph{93}, 032146\relax
\mciteBstWouldAddEndPuncttrue
\mciteSetBstMidEndSepPunct{\mcitedefaultmidpunct}
{\mcitedefaultendpunct}{\mcitedefaultseppunct}\relax
\EndOfBibitem
\bibitem[Saha \latin{et~al.}(2021)Saha, Lucero, Ehrich, Sivak, and Bechhoefer]{saha2021}
Saha,~T.~K.; Lucero,~J.~N.; Ehrich,~J.; Sivak,~D.~A.; Bechhoefer,~J. Maximizing power and velocity of an information engine. \emph{Proc. Natl. Acad. Sci.} \textbf{2021}, \emph{118}, e2023356118\relax
\mciteBstWouldAddEndPuncttrue
\mciteSetBstMidEndSepPunct{\mcitedefaultmidpunct}
{\mcitedefaultendpunct}{\mcitedefaultseppunct}\relax
\EndOfBibitem
\bibitem[Rafeek and Mondal(2024)Rafeek, and Mondal]{rafeek2024}
Rafeek,~R.; Mondal,~D. Active Brownian information engine: Self-propulsion induced colossal performance. \emph{J. Chem. Phys.} \textbf{2024}, \emph{161}, 124116\relax
\mciteBstWouldAddEndPuncttrue
\mciteSetBstMidEndSepPunct{\mcitedefaultmidpunct}
{\mcitedefaultendpunct}{\mcitedefaultseppunct}\relax
\EndOfBibitem
\bibitem[Paneru \latin{et~al.}(2022)Paneru, Dutta, and Pak]{Paneru_2022}
Paneru,~G.; Dutta,~S.; Pak,~H.~K. Colossal Power Extraction from Active Cyclic Brownian Information Engines. \emph{J. Phys. Chem. Lett.} \textbf{2022}, \emph{13}, 6912\relax
\mciteBstWouldAddEndPuncttrue
\mciteSetBstMidEndSepPunct{\mcitedefaultmidpunct}
{\mcitedefaultendpunct}{\mcitedefaultseppunct}\relax
\EndOfBibitem
\bibitem[Fang \latin{et~al.}(2019)Fang, Kruse, Lu, and Wang]{fang2019}
Fang,~X.; Kruse,~K.; Lu,~T.; Wang,~J. Nonequilibrium physics in biology. \emph{Rev. Mod. Phys.} \textbf{2019}, \emph{91}, 045004\relax
\mciteBstWouldAddEndPuncttrue
\mciteSetBstMidEndSepPunct{\mcitedefaultmidpunct}
{\mcitedefaultendpunct}{\mcitedefaultseppunct}\relax
\EndOfBibitem
\bibitem[Horowitz \latin{et~al.}(2013)Horowitz, Sagawa, and Parrondo]{Horowitz2013}
Horowitz,~J.~M.; Sagawa,~T.; Parrondo,~J. M.~R. Imitating Chemical Motors with Optimal Information Motors. \emph{Phys. Rev. Lett.} \textbf{2013}, \emph{111}, 010602\relax
\mciteBstWouldAddEndPuncttrue
\mciteSetBstMidEndSepPunct{\mcitedefaultmidpunct}
{\mcitedefaultendpunct}{\mcitedefaultseppunct}\relax
\EndOfBibitem
\bibitem[Linke and Parrondo(2021)Linke, and Parrondo]{Heiner_2021}
Linke,~H.; Parrondo,~J. M.~R. Tuning up Maxwell’s demon. \emph{Proc. Nat. Acad. Sci.} \textbf{2021}, \emph{118}, e2108218118\relax
\mciteBstWouldAddEndPuncttrue
\mciteSetBstMidEndSepPunct{\mcitedefaultmidpunct}
{\mcitedefaultendpunct}{\mcitedefaultseppunct}\relax
\EndOfBibitem
\bibitem[Borsley \latin{et~al.}(2021)Borsley, Leigh, and Roberts]{borsley_2021}
Borsley,~S.; Leigh,~D.~A.; Roberts,~B. M.~W. A Doubly Kinetically-Gated Information Ratchet Autonomously Driven by Carbodiimide Hydration. \emph{J. Am. Chem. Soc.} \textbf{2021}, \emph{143}, 4414, PMID: 33705123\relax
\mciteBstWouldAddEndPuncttrue
\mciteSetBstMidEndSepPunct{\mcitedefaultmidpunct}
{\mcitedefaultendpunct}{\mcitedefaultseppunct}\relax
\EndOfBibitem
\bibitem[Chakraborty \latin{et~al.}(2023)Chakraborty, Yang, Wang, and Zhong]{chakraborty2023}
Chakraborty,~D.; Yang,~C.; Wang,~L.; Zhong,~D. Role of Substrate Binding Interactions on DNA Repair by Photolyase. \emph{J. Phys. Chem. Lett.} \textbf{2023}, \emph{14}, 6672\relax
\mciteBstWouldAddEndPuncttrue
\mciteSetBstMidEndSepPunct{\mcitedefaultmidpunct}
{\mcitedefaultendpunct}{\mcitedefaultseppunct}\relax
\EndOfBibitem
\bibitem[Tomé and de~Oliveira(2018)Tomé, and de~Oliveira]{Tomé2018}
Tomé,~T.; de~Oliveira,~M.~J. Stochastic thermodynamics and entropy production of chemical reaction systems. \emph{J. Chem. Phys.} \textbf{2018}, \emph{148}, 224104\relax
\mciteBstWouldAddEndPuncttrue
\mciteSetBstMidEndSepPunct{\mcitedefaultmidpunct}
{\mcitedefaultendpunct}{\mcitedefaultseppunct}\relax
\EndOfBibitem
\bibitem[Schmiedl and Seifert(2007)Schmiedl, and Seifert]{Schmiedl2007}
Schmiedl,~T.; Seifert,~U. Stochastic thermodynamics of chemical reaction networks. \emph{J. Chem. Phys.} \textbf{2007}, \emph{126}, 044101\relax
\mciteBstWouldAddEndPuncttrue
\mciteSetBstMidEndSepPunct{\mcitedefaultmidpunct}
{\mcitedefaultendpunct}{\mcitedefaultseppunct}\relax
\EndOfBibitem
\bibitem[Wu and Wang(2014)Wu, and Wang]{wu2014}
Wu,~W.; Wang,~J. Potential and flux field landscape theory. II. Non-equilibrium thermodynamics of spatially inhomogeneous stochastic dynamical systems. \emph{J. Chem. Phys.} \textbf{2014}, \emph{141}, 105104\relax
\mciteBstWouldAddEndPuncttrue
\mciteSetBstMidEndSepPunct{\mcitedefaultmidpunct}
{\mcitedefaultendpunct}{\mcitedefaultseppunct}\relax
\EndOfBibitem
\bibitem[Xiao \latin{et~al.}(2009)Xiao, Hou, and Xin]{xiao2009}
Xiao,~T.; Hou,~Z.; Xin,~H. Stochastic Thermodynamics in Mesoscopic Chemical Oscillation Systems. \emph{J. Phys. Chem. B} \textbf{2009}, \emph{113}, 9316, PMID: 19518121\relax
\mciteBstWouldAddEndPuncttrue
\mciteSetBstMidEndSepPunct{\mcitedefaultmidpunct}
{\mcitedefaultendpunct}{\mcitedefaultseppunct}\relax
\EndOfBibitem
\bibitem[Fritz \latin{et~al.}(2020)Fritz, Nguyen, and Seifert]{Fritz2020}
Fritz,~J.~H.; Nguyen,~B.; Seifert,~U. Stochastic thermodynamics of chemical reactions coupled to finite reservoirs: A case study for the Brusselator. \emph{J. Chem. Phys.} \textbf{2020}, \emph{152}, 235101\relax
\mciteBstWouldAddEndPuncttrue
\mciteSetBstMidEndSepPunct{\mcitedefaultmidpunct}
{\mcitedefaultendpunct}{\mcitedefaultseppunct}\relax
\EndOfBibitem
\bibitem[Horowitz(2015)]{Horowitz2015}
Horowitz,~J.~M. Diffusion approximations to the chemical master equation only have a consistent stochastic thermodynamics at chemical equilibrium. \emph{J. Chem. Phys.} \textbf{2015}, \emph{143}, 044111\relax
\mciteBstWouldAddEndPuncttrue
\mciteSetBstMidEndSepPunct{\mcitedefaultmidpunct}
{\mcitedefaultendpunct}{\mcitedefaultseppunct}\relax
\EndOfBibitem
\bibitem[Mou \latin{et~al.}(1986)Mou, Luo, and Nicolis]{Mou1986}
Mou,~C.~Y.; Luo,~J.; Nicolis,~G. Stochastic thermodynamics of nonequilibrium steady states in chemical reaction systems. \emph{J. Chem. Phys.} \textbf{1986}, \emph{84}, 7011--7017\relax
\mciteBstWouldAddEndPuncttrue
\mciteSetBstMidEndSepPunct{\mcitedefaultmidpunct}
{\mcitedefaultendpunct}{\mcitedefaultseppunct}\relax
\EndOfBibitem
\bibitem[Barato \latin{et~al.}(2013)Barato, Hartich, and Seifert]{barato2013}
Barato,~A.~C.; Hartich,~D.; Seifert,~U. Information-theoretic versus thermodynamic entropy production in autonomous sensory networks. \emph{Phys. Rev. E} \textbf{2013}, \emph{87}, 042104\relax
\mciteBstWouldAddEndPuncttrue
\mciteSetBstMidEndSepPunct{\mcitedefaultmidpunct}
{\mcitedefaultendpunct}{\mcitedefaultseppunct}\relax
\EndOfBibitem
\bibitem[Barato and Seifert(2014)Barato, and Seifert]{barato2014}
Barato,~A.~C.; Seifert,~U. Stochastic thermodynamics with information reservoirs. \emph{Phys. Rev. E} \textbf{2014}, \emph{90}, 042150\relax
\mciteBstWouldAddEndPuncttrue
\mciteSetBstMidEndSepPunct{\mcitedefaultmidpunct}
{\mcitedefaultendpunct}{\mcitedefaultseppunct}\relax
\EndOfBibitem
\bibitem[Kim \latin{et~al.}(2011)Kim, Sagawa, De~Liberato, and Ueda]{Kim2011prl}
Kim,~S.~W.; Sagawa,~T.; De~Liberato,~S.; Ueda,~M. Quantum {S}zilard {E}ngine. \emph{Phys. Rev. Lett.} \textbf{2011}, \emph{106}, 070401\relax
\mciteBstWouldAddEndPuncttrue
\mciteSetBstMidEndSepPunct{\mcitedefaultmidpunct}
{\mcitedefaultendpunct}{\mcitedefaultseppunct}\relax
\EndOfBibitem
\bibitem[Bruschi \latin{et~al.}(2015)Bruschi, Perarnau-Llobet, Friis, Hovhannisyan, and Huber]{Bruschi2015pre}
Bruschi,~D.~E.; Perarnau-Llobet,~M.; Friis,~N.; Hovhannisyan,~K.~V.; Huber,~M. Thermodynamics of creating correlations: Limitations and optimal protocols. \emph{Phys. Rev. E} \textbf{2015}, \emph{91}, 032118\relax
\mciteBstWouldAddEndPuncttrue
\mciteSetBstMidEndSepPunct{\mcitedefaultmidpunct}
{\mcitedefaultendpunct}{\mcitedefaultseppunct}\relax
\EndOfBibitem
\bibitem[Goold \latin{et~al.}(2016)Goold, Huber, Riera, Del~Rio, and Skrzypczyk]{Goold2016jphysA}
Goold,~J.; Huber,~M.; Riera,~A.; Del~Rio,~L.; Skrzypczyk,~P. The role of quantum information in thermodynamics — a topical review. \emph{J. Phys. A} \textbf{2016}, \emph{49}, 143001\relax
\mciteBstWouldAddEndPuncttrue
\mciteSetBstMidEndSepPunct{\mcitedefaultmidpunct}
{\mcitedefaultendpunct}{\mcitedefaultseppunct}\relax
\EndOfBibitem
\bibitem[Koski \latin{et~al.}(2014)Koski, Maisi, Pekola, and Averin]{koski2014pnas}
Koski,~J.~V.; Maisi,~V.~F.; Pekola,~J.~P.; Averin,~D.~V. Experimental realization of a Szilard engine with a single electron. \emph{Proc. Natl. Acad. Sci.} \textbf{2014}, \emph{111}, 13786\relax
\mciteBstWouldAddEndPuncttrue
\mciteSetBstMidEndSepPunct{\mcitedefaultmidpunct}
{\mcitedefaultendpunct}{\mcitedefaultseppunct}\relax
\EndOfBibitem
\bibitem[Paneru \latin{et~al.}(2020)Paneru, Dutta, Sagawa, Tlusty, and Pak]{Paneru2020natcommun}
Paneru,~G.; Dutta,~S.; Sagawa,~T.; Tlusty,~T.; Pak,~H.~K. Efficiency fluctuations and noise induced refrigerator-to-heater transition in information engines. \emph{Nat. Commun.} \textbf{2020}, \emph{11}, 1012\relax
\mciteBstWouldAddEndPuncttrue
\mciteSetBstMidEndSepPunct{\mcitedefaultmidpunct}
{\mcitedefaultendpunct}{\mcitedefaultseppunct}\relax
\EndOfBibitem
\bibitem[Toyabe \latin{et~al.}(2010)Toyabe, Sagawa, Ueda, Muneyuki, and Sano]{Toyabe2010natphys}
Toyabe,~S.; Sagawa,~T.; Ueda,~M.; Muneyuki,~E.; Sano,~M. Experimental demonstration of information-to-energy conversion and validation of the generalized {J}arzynski equality. \emph{Nat. Phys.} \textbf{2010}, \emph{6}, 988--992\relax
\mciteBstWouldAddEndPuncttrue
\mciteSetBstMidEndSepPunct{\mcitedefaultmidpunct}
{\mcitedefaultendpunct}{\mcitedefaultseppunct}\relax
\EndOfBibitem
\bibitem[B{\'e}rut \latin{et~al.}(2012)B{\'e}rut, Arakelyan, Petrosyan, Ciliberto, Dillenschneider, and Lutz]{Berut2012nat}
B{\'e}rut,~A.; Arakelyan,~A.; Petrosyan,~A.; Ciliberto,~S.; Dillenschneider,~R.; Lutz,~E. Experimental verification of {L}andauer's principle linking information and thermodynamics. \emph{Nature} \textbf{2012}, \emph{483}, 187\relax
\mciteBstWouldAddEndPuncttrue
\mciteSetBstMidEndSepPunct{\mcitedefaultmidpunct}
{\mcitedefaultendpunct}{\mcitedefaultseppunct}\relax
\EndOfBibitem
\bibitem[Paneru \latin{et~al.}(2018)Paneru, Lee, Tlusty, and Pak]{Paneru2018prl}
Paneru,~G.; Lee,~D.~Y.; Tlusty,~T.; Pak,~H.~K. Lossless {B}rownian {I}nformation engine. \emph{Phys. Rev. Lett.} \textbf{2018}, \emph{120}, 020601\relax
\mciteBstWouldAddEndPuncttrue
\mciteSetBstMidEndSepPunct{\mcitedefaultmidpunct}
{\mcitedefaultendpunct}{\mcitedefaultseppunct}\relax
\EndOfBibitem
\bibitem[Lopez \latin{et~al.}(2008)Lopez, Kuwada, Craig, Long, and Linke]{Lopez2008prl}
Lopez,~B.~J.; Kuwada,~N.~J.; Craig,~E.~M.; Long,~B.~R.; Linke,~H. Realization of a Feedback Controlled Flashing Ratchet. \emph{Phys. Rev. Lett.} \textbf{2008}, \emph{101}, 220601\relax
\mciteBstWouldAddEndPuncttrue
\mciteSetBstMidEndSepPunct{\mcitedefaultmidpunct}
{\mcitedefaultendpunct}{\mcitedefaultseppunct}\relax
\EndOfBibitem
\bibitem[Dago \latin{et~al.}(2021)Dago, Pereda, Barros, Ciliberto, and Bellon]{dago2021}
Dago,~S.; Pereda,~J.; Barros,~N.; Ciliberto,~S.; Bellon,~L. Information and thermodynamics: Fast and precise approach to landauer’s bound in an underdamped micromechanical oscillator. \emph{Phys. Rev. Lett.} \textbf{2021}, \emph{126}, 170601\relax
\mciteBstWouldAddEndPuncttrue
\mciteSetBstMidEndSepPunct{\mcitedefaultmidpunct}
{\mcitedefaultendpunct}{\mcitedefaultseppunct}\relax
\EndOfBibitem
\bibitem[Dago and Bellon(2022)Dago, and Bellon]{dago2022}
Dago,~S.; Bellon,~L. Dynamics of Information Erasure and Extension of Landauer's Bound to Fast Processes. \emph{Phys. Rev. Lett.} \textbf{2022}, \emph{128}, 070604\relax
\mciteBstWouldAddEndPuncttrue
\mciteSetBstMidEndSepPunct{\mcitedefaultmidpunct}
{\mcitedefaultendpunct}{\mcitedefaultseppunct}\relax
\EndOfBibitem
\bibitem[Archambault \latin{et~al.}(2024)Archambault, Crauste-Thibierge, Ciliberto, and Bellon]{archambault2024}
Archambault,~A.; Crauste-Thibierge,~C.; Ciliberto,~S.; Bellon,~L. Inertial effects in discrete sampling information engines. \emph{Europhys. Lett.} \textbf{2024}, \emph{148}, 41002\relax
\mciteBstWouldAddEndPuncttrue
\mciteSetBstMidEndSepPunct{\mcitedefaultmidpunct}
{\mcitedefaultendpunct}{\mcitedefaultseppunct}\relax
\EndOfBibitem
\bibitem[Paneru \latin{et~al.}(2018)Paneru, Lee, Park, Park, Noh, and Pak]{paneru2018pre}
Paneru,~G.; Lee,~D.~Y.; Park,~J.-M.; Park,~J.~T.; Noh,~J.~D.; Pak,~H.~K. Optimal tuning of a Brownian information engine operating in a nonequilibrium steady state. \emph{Phys. Rev. E} \textbf{2018}, \emph{98}, 052119\relax
\mciteBstWouldAddEndPuncttrue
\mciteSetBstMidEndSepPunct{\mcitedefaultmidpunct}
{\mcitedefaultendpunct}{\mcitedefaultseppunct}\relax
\EndOfBibitem
\bibitem[Zwanzig(1992)]{zwanzig1992}
Zwanzig,~R. Diffusion past an entropy barrier. \emph{J. Phys. Chem.} \textbf{1992}, \emph{96}, 3926\relax
\mciteBstWouldAddEndPuncttrue
\mciteSetBstMidEndSepPunct{\mcitedefaultmidpunct}
{\mcitedefaultendpunct}{\mcitedefaultseppunct}\relax
\EndOfBibitem
\bibitem[Reguera and Rubi(2001)Reguera, and Rubi]{reguera2001}
Reguera,~D.; Rubi,~J. Kinetic equations for diffusion in the presence of entropic barriers. \emph{Phys. Rev. E} \textbf{2001}, \emph{64}, 061106\relax
\mciteBstWouldAddEndPuncttrue
\mciteSetBstMidEndSepPunct{\mcitedefaultmidpunct}
{\mcitedefaultendpunct}{\mcitedefaultseppunct}\relax
\EndOfBibitem
\bibitem[Mondal and Ray(2010)Mondal, and Ray]{mondal2010diffusion}
Mondal,~D.; Ray,~D.~S. Diffusion over an entropic barrier: Non-Arrhenius behavior. \emph{Phys. Rev. E} \textbf{2010}, \emph{82}, 032103\relax
\mciteBstWouldAddEndPuncttrue
\mciteSetBstMidEndSepPunct{\mcitedefaultmidpunct}
{\mcitedefaultendpunct}{\mcitedefaultseppunct}\relax
\EndOfBibitem
\bibitem[Ghosh \latin{et~al.}(2010)Ghosh, Marchesoni, Savel’ev, and Nori]{ghosh2010}
Ghosh,~P.~K.; Marchesoni,~F.; Savel’ev,~S.~E.; Nori,~F. Geometric stochastic resonance. \emph{Phys. Rev. Lett.} \textbf{2010}, \emph{104}, 020601\relax
\mciteBstWouldAddEndPuncttrue
\mciteSetBstMidEndSepPunct{\mcitedefaultmidpunct}
{\mcitedefaultendpunct}{\mcitedefaultseppunct}\relax
\EndOfBibitem
\bibitem[Ali \latin{et~al.}(2024)Ali, Bauri, and Mondal]{ali2024}
Ali,~S.~Y.; Bauri,~P.; Mondal,~D. Optimizing work extraction in the presence of an entropic potential: An entropic stochastic resonance. \emph{J. Phys. Chem. B} \textbf{2024}, \emph{128}, 3824--3832\relax
\mciteBstWouldAddEndPuncttrue
\mciteSetBstMidEndSepPunct{\mcitedefaultmidpunct}
{\mcitedefaultendpunct}{\mcitedefaultseppunct}\relax
\EndOfBibitem
\bibitem[Chupeau \latin{et~al.}(2020)Chupeau, Gladrow, Chepelianskii, Keyser, and Trizac]{marie2020}
Chupeau,~M.; Gladrow,~J.; Chepelianskii,~A.; Keyser,~U.~F.; Trizac,~E. Optimizing Brownian escape rates by potential shaping. \emph{Proc. Natl. Acad. Sci.} \textbf{2020}, \emph{117}, 1383\relax
\mciteBstWouldAddEndPuncttrue
\mciteSetBstMidEndSepPunct{\mcitedefaultmidpunct}
{\mcitedefaultendpunct}{\mcitedefaultseppunct}\relax
\EndOfBibitem
\bibitem[Zwanzig(1988)]{zwanzig1988}
Zwanzig,~R. Diffusion in a rough potential. \emph{Proc. Natl. Acad. Sci.} \textbf{1988}, \emph{85}, 2029\relax
\mciteBstWouldAddEndPuncttrue
\mciteSetBstMidEndSepPunct{\mcitedefaultmidpunct}
{\mcitedefaultendpunct}{\mcitedefaultseppunct}\relax
\EndOfBibitem
\bibitem[Mondal(2011)]{mondal2011}
Mondal,~D. Enhancement of entropic transport by intermediates. \emph{Phys. Rev. E} \textbf{2011}, \emph{84}, 011149\relax
\mciteBstWouldAddEndPuncttrue
\mciteSetBstMidEndSepPunct{\mcitedefaultmidpunct}
{\mcitedefaultendpunct}{\mcitedefaultseppunct}\relax
\EndOfBibitem
\bibitem[Ghosh \latin{et~al.}(2007)Ghosh, Bag, and Ray]{ghosh2007}
Ghosh,~P.~K.; Bag,~B.~C.; Ray,~D.~S. {Noise correlation-induced splitting of Kramers’ escape rate from a metastable state}. \emph{J. Chem. Phys.} \textbf{2007}, \emph{127}, 044510\relax
\mciteBstWouldAddEndPuncttrue
\mciteSetBstMidEndSepPunct{\mcitedefaultmidpunct}
{\mcitedefaultendpunct}{\mcitedefaultseppunct}\relax
\EndOfBibitem
\bibitem[Mondal \latin{et~al.}(2010)Mondal, Das, and Ray]{mondal2010entropic}
Mondal,~D.; Das,~M.; Ray,~D.~S. Entropic noise-induced nonequilibrium transition. \emph{J. Chem. Phys.} \textbf{2010}, \emph{133}\relax
\mciteBstWouldAddEndPuncttrue
\mciteSetBstMidEndSepPunct{\mcitedefaultmidpunct}
{\mcitedefaultendpunct}{\mcitedefaultseppunct}\relax
\EndOfBibitem
\bibitem[McNamara and Wiesenfeld(1989)McNamara, and Wiesenfeld]{mcnamara1989}
McNamara,~B.; Wiesenfeld,~K. Theory of stochastic resonance. \emph{Phys. Rev. A} \textbf{1989}, \emph{39}, 4854\relax
\mciteBstWouldAddEndPuncttrue
\mciteSetBstMidEndSepPunct{\mcitedefaultmidpunct}
{\mcitedefaultendpunct}{\mcitedefaultseppunct}\relax
\EndOfBibitem
\bibitem[Mondal and Muthukumar(2016)Mondal, and Muthukumar]{mondal2016resonance}
Mondal,~D.; Muthukumar,~M. Stochastic resonance during a polymer translocation process. \emph{J. Chem. Phys.} \textbf{2016}, \emph{144}, 144901\relax
\mciteBstWouldAddEndPuncttrue
\mciteSetBstMidEndSepPunct{\mcitedefaultmidpunct}
{\mcitedefaultendpunct}{\mcitedefaultseppunct}\relax
\EndOfBibitem
\bibitem[Zhu \latin{et~al.}(2022)Zhu, Zhou, Marchesoni, and Zhang]{zhu2022}
Zhu,~Q.; Zhou,~Y.; Marchesoni,~F.; Zhang,~H.~P. Colloidal Stochastic Resonance in Confined Geometries. \emph{Phys. Rev. Lett.} \textbf{2022}, \emph{129}, 098001\relax
\mciteBstWouldAddEndPuncttrue
\mciteSetBstMidEndSepPunct{\mcitedefaultmidpunct}
{\mcitedefaultendpunct}{\mcitedefaultseppunct}\relax
\EndOfBibitem
\bibitem[Mondal \latin{et~al.}(2012)Mondal, Das, and Ray]{mondal2012entropic}
Mondal,~D.; Das,~M.; Ray,~D.~S. Entropic dynamical hysteresis in a driven system. \emph{Phys. Rev. E} \textbf{2012}, \emph{85}, 031128\relax
\mciteBstWouldAddEndPuncttrue
\mciteSetBstMidEndSepPunct{\mcitedefaultmidpunct}
{\mcitedefaultendpunct}{\mcitedefaultseppunct}\relax
\EndOfBibitem
\bibitem[Doering and Gadoua(1992)Doering, and Gadoua]{doering1992}
Doering,~C.~R.; Gadoua,~J.~C. Resonant activation over a fluctuating barrier. \emph{Phys. Rev. Lett.} \textbf{1992}, \emph{69}, 2318\relax
\mciteBstWouldAddEndPuncttrue
\mciteSetBstMidEndSepPunct{\mcitedefaultmidpunct}
{\mcitedefaultendpunct}{\mcitedefaultseppunct}\relax
\EndOfBibitem
\bibitem[Reimann(1995)]{reimann1995}
Reimann,~P. Thermally Driven Escape with Fluctuating Potentials: A New Type of Resonant Activation. \emph{Phys. Rev. Lett.} \textbf{1995}, \emph{74}, 4576\relax
\mciteBstWouldAddEndPuncttrue
\mciteSetBstMidEndSepPunct{\mcitedefaultmidpunct}
{\mcitedefaultendpunct}{\mcitedefaultseppunct}\relax
\EndOfBibitem
\bibitem[Iwaniszewski(1996)]{jan1996}
Iwaniszewski,~J. Escape over a fluctuating barrier: Limits of small and large correlation times. \emph{Phys. Rev. E} \textbf{1996}, \emph{54}, 3173\relax
\mciteBstWouldAddEndPuncttrue
\mciteSetBstMidEndSepPunct{\mcitedefaultmidpunct}
{\mcitedefaultendpunct}{\mcitedefaultseppunct}\relax
\EndOfBibitem
\bibitem[Mondal \latin{et~al.}(2010)Mondal, Das, and Ray]{mondal2010resonant}
Mondal,~D.; Das,~M.; Ray,~D.~S. Entropic resonant activation. \emph{J. Chem. Phys.} \textbf{2010}, \emph{132}\relax
\mciteBstWouldAddEndPuncttrue
\mciteSetBstMidEndSepPunct{\mcitedefaultmidpunct}
{\mcitedefaultendpunct}{\mcitedefaultseppunct}\relax
\EndOfBibitem
\bibitem[Chattoraj \latin{et~al.}(2014)Chattoraj, Saha, Jana, and Bhattacharyya]{chattoraj2014dynamics}
Chattoraj,~S.; Saha,~S.; Jana,~S.~S.; Bhattacharyya,~K. Dynamics of gene silencing in a live cell: stochastic resonance. \emph{J. Phys. Chem. Lett.} \textbf{2014}, \emph{5}, 1012\relax
\mciteBstWouldAddEndPuncttrue
\mciteSetBstMidEndSepPunct{\mcitedefaultmidpunct}
{\mcitedefaultendpunct}{\mcitedefaultseppunct}\relax
\EndOfBibitem
\bibitem[Magnasco(1993)]{magnasco1993}
Magnasco,~M.~O. Forced thermal ratchets. \emph{Phys. Rev. Lett.} \textbf{1993}, \emph{71}, 1477\relax
\mciteBstWouldAddEndPuncttrue
\mciteSetBstMidEndSepPunct{\mcitedefaultmidpunct}
{\mcitedefaultendpunct}{\mcitedefaultseppunct}\relax
\EndOfBibitem
\bibitem[Mondal \latin{et~al.}(2009)Mondal, Ghosh, and Ray]{mondal2009roughratchet}
Mondal,~D.; Ghosh,~P.~K.; Ray,~D.~S. Noise-induced transport in a rough ratchet potential. \emph{J Chem. Phys.} \textbf{2009}, \emph{130}\relax
\mciteBstWouldAddEndPuncttrue
\mciteSetBstMidEndSepPunct{\mcitedefaultmidpunct}
{\mcitedefaultendpunct}{\mcitedefaultseppunct}\relax
\EndOfBibitem
\bibitem[Mondal and Muthukumar(2016)Mondal, and Muthukumar]{mondal2016ratchet}
Mondal,~D.; Muthukumar,~M. Ratchet rectification effect on the translocation of a flexible polyelectrolyte chain. \emph{J. Chem. Phys.} \textbf{2016}, \emph{145}, 084906\relax
\mciteBstWouldAddEndPuncttrue
\mciteSetBstMidEndSepPunct{\mcitedefaultmidpunct}
{\mcitedefaultendpunct}{\mcitedefaultseppunct}\relax
\EndOfBibitem
\bibitem[Kato and Tanimura(2013)Kato, and Tanimura]{kato2013quantum}
Kato,~A.; Tanimura,~Y. Quantum suppression of ratchet rectification in a Brownian system driven by a biharmonic force. \emph{J. Phys. Chem. B} \textbf{2013}, \emph{117}, 13132\relax
\mciteBstWouldAddEndPuncttrue
\mciteSetBstMidEndSepPunct{\mcitedefaultmidpunct}
{\mcitedefaultendpunct}{\mcitedefaultseppunct}\relax
\EndOfBibitem
\bibitem[Wagoner and Dill(2016)Wagoner, and Dill]{wagoner2016}
Wagoner,~J.~A.; Dill,~K.~A. Molecular Motors: Power Strokes Outperform Brownian Ratchets. \emph{J. Phys. Chem. B} \textbf{2016}, \emph{120}, 6327, PMID: 27136319\relax
\mciteBstWouldAddEndPuncttrue
\mciteSetBstMidEndSepPunct{\mcitedefaultmidpunct}
{\mcitedefaultendpunct}{\mcitedefaultseppunct}\relax
\EndOfBibitem
\bibitem[Rafeek and Mondal(2025)Rafeek, and Mondal]{rafna_pot_2025}
Rafeek,~R.; Mondal,~D. Achievable Information-Energy Exchange in a Brownian Information Engine through Potential Profiling. \emph{J. Phys. Chem. B} \textbf{2025}, \emph{129}, 2971\relax
\mciteBstWouldAddEndPuncttrue
\mciteSetBstMidEndSepPunct{\mcitedefaultmidpunct}
{\mcitedefaultendpunct}{\mcitedefaultseppunct}\relax
\EndOfBibitem
\bibitem[Risken(1996)]{Risken}
Risken,~H. \emph{The Fokker-Planck Equation}; Springer-Verlag Berlin Heidelberg, 1996\relax
\mciteBstWouldAddEndPuncttrue
\mciteSetBstMidEndSepPunct{\mcitedefaultmidpunct}
{\mcitedefaultendpunct}{\mcitedefaultseppunct}\relax
\EndOfBibitem
\bibitem[Van~Kampen(1992)]{van1992stochastic}
Van~Kampen,~N.~G. \emph{Stochastic processes in physics and chemistry}; Elsevier, 1992\relax
\mciteBstWouldAddEndPuncttrue
\mciteSetBstMidEndSepPunct{\mcitedefaultmidpunct}
{\mcitedefaultendpunct}{\mcitedefaultseppunct}\relax
\EndOfBibitem
\bibitem[Hildebrand(1987)]{Hildebrandnumerical}
Hildebrand,~F.~B. \emph{Introduction to {N}umerical {A}nalysis}; Courier Corporation, 1987\relax
\mciteBstWouldAddEndPuncttrue
\mciteSetBstMidEndSepPunct{\mcitedefaultmidpunct}
{\mcitedefaultendpunct}{\mcitedefaultseppunct}\relax
\EndOfBibitem
\bibitem[Box and Muller(1958)Box, and Muller]{box1958}
Box,~G.~E.; Muller,~M.~E. A note on the generation of random normal deviates. \emph{Ann. Math. Stat.} \textbf{1958}, \emph{29}, 610\relax
\mciteBstWouldAddEndPuncttrue
\mciteSetBstMidEndSepPunct{\mcitedefaultmidpunct}
{\mcitedefaultendpunct}{\mcitedefaultseppunct}\relax
\EndOfBibitem
\bibitem[Ramesh \latin{et~al.}(2025)Ramesh, Busink, Moesbergen, Peters, Ackermans, and Rodriguez]{Ramesh2025}
Ramesh,~V.~G.; Busink,~J.; Moesbergen,~R. E.~R.; Peters,~K. J.~H.; Ackermans,~P.~J.; Rodriguez,~S. R.~K. Stochastic Thermodynamics of a Linear Optical Cavity Driven on Resonance. \emph{ACS Photonics} \textbf{2025}, \emph{12}, 159\relax
\mciteBstWouldAddEndPuncttrue
\mciteSetBstMidEndSepPunct{\mcitedefaultmidpunct}
{\mcitedefaultendpunct}{\mcitedefaultseppunct}\relax
\EndOfBibitem
\bibitem[Kim and Hyeon(2021)Kim, and Hyeon]{Kim2021}
Kim,~P.; Hyeon,~C. Thermodynamic Optimality of Glycolytic Oscillations. \emph{J. Phys. Chem. B} \textbf{2021}, \emph{125}, 5740, PMID: 34038120\relax
\mciteBstWouldAddEndPuncttrue
\mciteSetBstMidEndSepPunct{\mcitedefaultmidpunct}
{\mcitedefaultendpunct}{\mcitedefaultseppunct}\relax
\EndOfBibitem
\bibitem[Cao and Hou(2022)Cao, and Hou]{Cao2022}
Cao,~Z.; Hou,~Z. Improved estimation for energy dissipation in biochemical oscillations. \emph{J. Chem. Phys.} \textbf{2022}, \emph{157}, 025102\relax
\mciteBstWouldAddEndPuncttrue
\mciteSetBstMidEndSepPunct{\mcitedefaultmidpunct}
{\mcitedefaultendpunct}{\mcitedefaultseppunct}\relax
\EndOfBibitem
\bibitem[Jop \latin{et~al.}(2008)Jop, Petrosyan, and Ciliberto]{jop2008work}
Jop,~P.; Petrosyan,~A.; Ciliberto,~S. Work and dissipation fluctuations near the stochastic resonance of a colloidal particle. \emph{Europhys. Lett.} \textbf{2008}, \emph{81}, 50005\relax
\mciteBstWouldAddEndPuncttrue
\mciteSetBstMidEndSepPunct{\mcitedefaultmidpunct}
{\mcitedefaultendpunct}{\mcitedefaultseppunct}\relax
\EndOfBibitem
\end{mcitethebibliography}

\end{document}